\documentclass[
onecolumn, %can be preprint or twocolumn
12pt, %font size; default is 10pt
tightenlines,
superscriptaddress, %addresses listed after each author name or after all names
nofootinbib, %footnotes go in bibliography or bottom of page
preprintnumbers, %show preprint numbers 
notitlepage, %remove page break after abstract
prd %style can be aps or prd or prl
]{revtex4-1}
\pdfoutput=1

\usepackage[centering,hmargin=2.5cm,vmargin=2.5cm]{geometry}
\usepackage[colorlinks=true,citecolor=blue,linkcolor=blue]{hyperref}
\usepackage[T1]{fontenc}
\usepackage[normalem]{ulem}
\usepackage{amsmath,amssymb}
\usepackage{mathtools}
\usepackage{epsfig}
\usepackage{graphicx} % Standard graphics package
\usepackage{grffile} % fix problem with graphics files whose names contain dots
\usepackage{url}
\usepackage{color}
\usepackage{multirow}
\usepackage{placeins}
\usepackage[dvipsnames]{xcolor}
\usepackage{epstopdf}
\usepackage{siunitx}
\usepackage{enumitem}
\usepackage{cleveref}
\usepackage{lipsum}
\usepackage{gensymb}
\usepackage{xspace}
\usepackage{titlesec} % modify titles and heading format
\usepackage{booktabs} % nicer tables
\usepackage{subfigure}
\usepackage[subfigure,titles]{tocloft} % subfigures,titles options needed to work with RevTeX

\allowdisplaybreaks

\makeatletter

\renewcommand{\p@subsection}{}

\renewcommand{\p@subsubsection}{}
\makeatother

\titleformat*{\section}{\centering\bfseries\scshape}
\titlelabel{\thetitle\quad}
\titleformat*{\paragraph}{\bfseries}
\titlespacing*{\paragraph}{0pt}{3.25ex plus 1ex minus .2ex}{1em}

\pacs{}
\keywords{}

\begin{document}

%=============================================================================
\title{Black Hole Evaporation Beyond the Standard Model of Particle Physics}

\author{Michael J.\ Baker}
\email{michael.baker@unimelb.edu.au}
\affiliation{ARC Centre of Excellence for Dark Matter Particle Physics, 
School of Physics, The University of Melbourne, Victoria 3010, Australia}

\author{Andrea Thamm}
\email{andrea.thamm@unimelb.edu.au}
\affiliation{ARC Centre of Excellence for Dark Matter Particle Physics, 
School of Physics, The University of Melbourne, Victoria 3010, Australia}

\date{\today}

%\preprint{}

%=============================================================================

\begin{abstract}
    \noindent
    The observation of an evaporating black hole would provide definitive information on the elementary particles present in nature.  In particular, it could discover or exclude particles beyond those present in the standard model of particle physics.  We consider a wide range of motivated scenarios beyond the standard model and identify those which would be best probed in the event of an observation.  For those models we define representative benchmark parameters and characterise the photon spectra as a function of time.  For the supersymmetric benchmark model, where most of the new particles produce secondary photons, we provide secondary spectra and discuss the subtle interplay between faster black hole evaporation and an increased flux of secondary photons.  Finally, we discuss the impact of these models on future experimental analysis strategies.
\end{abstract}

\maketitle

%=============================================================================
\section{Introduction}
\label{sec:introduction}
%=============================================================================

Black holes are now known to exist, and their properties are being studied via gravitational waves from binary black hole mergers and directly via photons emitted from their accretion disks.
Beyond these solar mass and supermassive black holes, much lighter black holes could exist in the universe.  For example, primordial black holes may have been produced in the early universe~\cite{
Carr:1975qj,
Ivanov:1994pa,
GarciaBellido:1996qt,
Silk:1986vc,
Kawasaki:1997ju,
Yokoyama:1995ex,
Pi:2017gih,
Hawking:1987bn,
Polnarev:1988dh,
MacGibbon:1997pu,
Rubin:2000dq,
Rubin:2001yw,
Brandenberger:2021zvn,
Cotner:2016cvr,
Cotner:2019ykd,
Crawford:1982yz,
Kodama:1982sf,
Moss:1994pi,
Freivogel:2007fx,
Hawking:1982ga,
Johnson:2011wt,
Cotner:2018vug,
Kusenko:2020pcg,
Baker:2021nyl,
Baker:2021sno}.  Those with masses around $10^{15}\,\text{g}$ would have been continually losing mass via Hawking radiation and would be reaching their final stages of evaporation today (see, e.g., refs.~\cite{Carr:2016drx,Sasaki:2018dmp,Green:2020jor,Carr:2020xqk,Carr:2020gox,Villanueva-Domingo:2021spv} for recent reviews of primordial black holes).  There are currently a range of experiments searching for both the explosive final stages of black hole evaporation, e.g.,~ref.~\cite{Albert:2019qxd}, and indirect signatures from a population of low-mass black holes, e.g.,~refs.~\cite{1976ApJ...206....1P,1977ApJ...212..224P,Lehoucq:2009ge,Abe:2011nx,Carr:2020gox}.

In this work we emphasise that the observation of an exploding black hole would provide definitive information on the particles present in nature, and that this in turn could provide evidence for, or rule out, a wide range of models Beyond the Standard Model (BSM).  A broad range of BSM models have been proposed and are widely studied.  These models usually aim to address a question not answered by the Standard Model (SM), such as the gauge hierarchy problem, the nature of dark matter, the strong-$CP$ problem, etc.  We first survey contemporary BSM models and identify those which are widely studied and would have a significant and calculable impact on the signal seen from a black hole explosion.  These are supersymmetry (for a recent review see, e.g., ref.~\cite{Workman:2022ynf}), $N$naturalness \cite{Arkani-Hamed:2016rle}, models inspired by string theory (for a recent review, see, e.g.,~ref.~\cite{Cvetic:2022fnv}) and dark sectors (e.g., refs.~\cite{Kobzarev:1966qya,Blinnikov:1982eh,Foot:1991bp,Hodges:1993yb,Berezhiani:1995am,Strassler:2006im,Cvetic:2012kj,Foot:2014uba}).  We then define representative benchmark parameter points for these models and determine the relevant particle mass spectra and decay properties.  This allows us to compute the mass evolution of the black hole and the primary and secondary photon spectra.  We use this to characterise the range of behaviours seen in this set of BSM models, and briefly discuss experimental strategies to distinguish between the SM and BSM scenarios. 

Conversely, we emphasise that assumptions about the particles present in nature will impact experimental limits on the rate-density of exploding black holes.  While searches for evaporating black holes typically assume the SM particles, we demonstrate that BSM models can dramatically alter the expected black hole evolution and the associated photon signatures. This means that if a BSM model is realised in nature, then the  direct and indirect limits could be drastically altered.  Similarly, existing gamma-ray bursts of unknown origin could potentially be attributed to evaporating black holes.  These may currently be missed in searches that assume only the SM particles.  The benchmark models we propose could therefore be used as a framework for a wider interpretation in future searches.

Previous work on black holes and BSM physics has mainly focused on BSM particle production in the early universe, e.g., refs.~\cite{hep-ph/9810456, astro-ph/9812301, astro-ph/9903484, hep-ph/0001238, astro-ph/0406621, Doran:2005mf, 0801.0116, 1401.1909, 1712.07664,  Johnson:2018gjr, 1812.10606, 1905.01301, 1910.07864, 2004.00618, 2004.04740, 2004.14773, Johnson:2020tiw, 2008.06505,  2010.01134, 2012.09867}. 
Ref.~\cite{Ukwatta:2015iba} considers the impact of a single $5\,\text{TeV}$ squark on the observation of an evaporating black hole and suggests that a deviation from the SM could be seen if a black hole explodes within $\sim 0.015\,\text{pc}$. 
In ref.~\cite{Baker:2021btk} we demonstrated that if the HAWC observatory observed $\sim 200$ photons from an exploding black hole it could probe dark sector models containing one or more copies of the SM particles with any mass scale up to $100\,\text{TeV}$.

%=============================================================================
\section{Beyond the Standard Model of Particle Physics}
\label{sec:models}
%=============================================================================

There are a wide variety of motivations for proposing new fundamental degrees of freedom (dof) beyond those present in the standard model of particle physics.  The standard model was chiefly developed and verified using particle colliders, and this is the environment where it is most applicable.  As such, the SM can not describe gravity, dark matter, dark energy, neutrino masses, or the matter-antimatter asymmetry.  There are also theoretical problems within the SM.  Those which have been most widely studied are the gauge hierarchy problem (why the Higgs boson has a mass at the weak scale and not at a higher scale) and the strong-$CP$ problem (which is related to the absence of $CP$ violation in the strong sector).  Furthermore, the SM has many unexplained features (such as three gauge groups, three generations, etc) and many unrelated parameters, which motivate the search for a deeper unified model.  There are also experimental results which are in tension with SM predictions, such as the flavour anomalies in $B$ physics and the anomalous magnetic moment of the muon.

To guide our identification of the most promising models to probe via black hole evaporation, we recap some of the conclusions from refs.~\cite{Ukwatta:2015iba,Baker:2021btk}.  Both ref.~\cite{Ukwatta:2015iba} and ref.~\cite{Baker:2021btk} demonstrate potential sensitivity to models of new physics.  One of the significant differences between the studies is that while the single squark analysed in ref.~\cite{Ukwatta:2015iba} produces extra secondary photons, the models considered in ref.~\cite{Baker:2021btk} do not.  However, even without extra secondary photons, `dark' degrees of freedom still take energy from the evaporating black hole and lead to an increased rate of mass loss, which is detectable in the photon signal originating from purely SM processes.  As such, it is not essential to consider only models that produce extra photons.  We also emphasise that Hawking radiation is independent of all non-gravitational couplings.  As such, the increased rate of mass loss induced by new degrees of freedom depends only on the particle mass and spin, and an observation can probe models that are essentially decoupled from the SM.

The BSM models that will have the largest impact on the signal from an evaporating black hole are those with a large number of new degrees of freedom, at a mass scale that is not too high. For example, ref.~\cite{Baker:2021btk} shows that the observation of 200 photons from an evaporating black hole at the HAWC observatory would be able to probe a dark sector containing one copy of the SM, around 100 new degrees of freedom, at any mass scale below $10^5\,\text{GeV}$.  The observation of 10 photons would be enough to probe ten copies of the SM (around 1000 new dof) up to a similar mass scale.  This provides an initial indication that we should first focus on models with $\gtrsim 100$ new degrees of freedom.

We will also focus on models that have standard black hole evaporation at black hole temperatures below $\sim10^7\,\text{GeV}$.  Models with a fundamental Planck scale below $\sim10^7\,\text{GeV}$ would lead to a striking signature where the black hole evaporation suddenly stops when the fundamental Planck scale is reached.  This would be the case in the large $N$ species and extra-dimensional models discussed below.  However, modelling of the final burst requires some assumptions about the effects of quantum gravity.  Furthermore, some BSM scenarios, such as extra-dimensional models, modify black hole dynamics below the fundamental Planck scale.  These scenarios require detailed, model-specific study which we defer to future work.  Ref.~\cite{Baker:2021btk} demonstrates that it is unlikely that searches will be sensitive to physics above $\sim10^7\,\text{GeV}$, so we will still consider BSM scenarios which modify black hole dynamics above this scale (such as $N$naturalness).

Astrophysical observations will typically only be sensitive to photons above a certain energy cutoff.  HAWC, for example, only has a significant effective area at $E_\gamma \gtrsim 10^2\,\text{GeV}$.  Furthermore, an analysis may only want to select higher energy events for effective background reduction.  This means that while these observations will be able to infer the presence of new dof with masses below these energies, it will not be sensitive to their precise mass scales.  While this is a reasonable cutoff for HAWC-like experiments, where an exploding black hole would likely first be seen, a lower cutoff may be more appropriate for exploding black holes seen using other experimental techniques or for attempts to probe BSM models using an integrated flux of lower energy photons from a more distant population of evaporating black holes.  For this reason, we consider new particles to be `massless' if, for a black hole exploding today, they could have been produced by the black hole shortly after the Big Bang.

We now survey contemporary models in BSM physics, with an emphasis on those that are widely studied and/or are expected to have a significant impact on the photon signal from a nearby evaporating black hole:
\begin{itemize}
    
    \item Supersymmetry -- From the 1980's to the mid-2010's supersymmetry was very widely studied as it could address the gauge hierarchy problem, gauge coupling unification and dark matter, it is a necessary ingredient of string theory, and it was widely expected to lead to new TeV scale particles (for a recent review, see, e.g., ref.~\cite{Workman:2022ynf}).  Although these particles have not been seen at the LHC or in dark matter direct detection experiments, supersymmetry is still studied and certain regions of parameter space remain viable.  In contrast to the other models we highlight, most of these new particles will produce secondary photons.

    \item Large $N$ Species Solution to the Hierarchy Problem~\cite{Dvali:2007hz,Dvali:2007wp,Calmet:2008tn,Dvali:2009ne} and $N$naturalness~\cite{Arkani-Hamed:2016rle} --   These models relax the hierarchy problem since the apparent Planck mass $M_\text{Pl}$ is related to the scale at which gravity becomes strongly coupled, $M_\ast$, by $M_\text{Pl}^2 \gtrsim N M_\ast^2$. In refs.~\cite{Dvali:2007hz,Dvali:2007wp,Calmet:2008tn,Dvali:2009ne} the hierarchy problem can be solved when gravity becomes strongly coupled at the TeV scale, which can be achieved for $N \sim 10^{32}$ copies of the SM.  However, this would mean that black hole evaporation stops at the TeV scale, rather than continuing above $10^7\,\text{GeV}$, so we do not consider this scenario in this work.  $N$naturalness~\cite{Arkani-Hamed:2016rle} solves the hierarchy problem while introducing fewer copies, and can retain a Planck scale above $10^7\,\text{GeV}$.  This is the scenario that we will focus on.
    
    \item String Inspired Models -- String theory is the most promising approach for combining general relativity and quantum field theory, to provide a quantum theory of gravity (for a recent review, see, e.g.,~ref.~\cite{Cvetic:2022fnv}).  In string theory, the fundamental particles of quantum field theory are replaced by fundamental one-dimensional strings.  While this class of theories is not yet well understood and a realistic model (in the sense of containing the SM particles) is yet to be constructed, some properties of the theory are relatively well understood.  One generic prediction is an abundance of particles, called \textit{moduli}, which are naively very light and which must obtain a mass through some mechanism to satisfy cosmological constraints.  These moduli may be light enough to impact the signal from an evaporating black hole, and we study two `string inspired' scenarios containing these moduli fields.
    
    \item Dark Sectors and Hidden Valleys -- The presence of dark matter is perhaps the strongest direct evidence for new particles beyond the SM.  The simplest and most widely studied models introduce just a few new degrees of freedom, and as such have a relatively small impact on the signal from an evaporating black hole.  However, dark matter could be part of a richer dark sector containing multiple new particles and dark gauge forces.  Hidden valley models~\cite{Strassler:2006im} similarly introduce a rich sector that only weakly communicates with the SM.  A representative class of dark sector models was studied in ref.~\cite{Baker:2021btk} and we include these models here for comparison.

    \item Extra-Dimensional Models -- These models were initially introduced to address the gauge hierarchy problem and are widely studied alternatives to supersymmetry.  In contrast to the other models we consider, extra-dimensional models alter the space-time geometry and so fundamentally change the black holes themselves, making the situation more complicated.  The main classes of models are large extra-dimensional models~\cite{Kaluza:1921tu,Klein:1926tv,Arkani-Hamed:1998jmv,Antoniadis:1998ig} and warped extra-dimensional models~\cite{Rubakov:1983bz,Rubakov:1983bb,Randall:1999ee,Randall:1999vf}.  Large extra-dimensional models lower the Planck scale, so these models suggest that black holes could be produced at colliders~\cite{Dimopoulos:2001hw,Cavaglia:2003hg} or in cosmic ray collisions~\cite{Feng:2001ib,Anchordoqui:2001ei,Emparan:2001kf,Ringwald:2001vk} (although none have yet been observed).  While this would significantly alter the signature, since the black hole would reach the end-point of its evaporation at a temperature $\ll 10^{18}\,\text{GeV}$, it would likely do so in a way that depends on the details of quantum gravity \cite{Harris:2003eg,Kanti:2004nr,Carr:2004kgc,Cardoso:2005vb,Cardoso:2005mh,Draggiotis:2008jz,Kanti:2014vsa,Johnson:2020tiw}.  While this may produce a signature quite obviously different from the SM expectation, quantitative analysis requires detailed study which we leave to future work. 
    While warped extra-dimensional models maintain a larger Planck scale~\cite{Randall:1999ee,Randall:1999vf,Csaki:2004ay}, no analytical solution exists for a five-dimensional asymptotically anti-de Sitter black hole localised on the brane and therefore the impact of the extra-dimension on the black hole evaporation is not well understood~\cite{Kanti:2014vsa}.
    As such we also do not consider these models in this work.
    
\end{itemize}

There are a host of other motivated models which could potentially be probed through the observation of an evaporating black hole.  However, since they require dedicated study or will only produce small deviations from the SM signal, we do not propose benchmarks for:
\begin{itemize}
    
    \item Composite Higgs Models -- These models also address the gauge hierarchy problem and are widely studied alternatives to supersymmetry and extra-dimensional models.  Although warped extra-dimensional models are in some ways dual to a wide class of composite Higgs models, this is not true for black hole evaporation.  While this could provide an intriguing way of distinguishing between these models in the event of more conventional evidence for these theories, confinement in the models makes the situation very complicated.  In the same way that Hawking evaporation near the QCD scale is beset with difficulties which are debated in the literature~\cite{MacGibbon:1990zk,MacGibbon:1991tj,Coogan:2020tuf,Arbey:2021mbl}, similar problems would emerge at the new confinement scale. We also do not expect a very large number of fundamental degrees of freedom in these theories, so we do not expect their impact to be particularly large.
    
    \item Grand Unified Theories -- While GUT theories are strongly motivated and aim to unify the gauge forces into one structure, they typically have new degrees of freedom at around $10^{16}\,\text{GeV}$, the expected unification scale.   Since this scale is so high, these models will only impact an evaporating black hole in the very last moments of its life and so will be very hard to probe via black hole explosion.  
    
    \item Light Dark Matter, Axions, ALPs, Dark Photons -- Despite an intensive, dedicated search for WIMP dark matter, no convincing signal has yet been observed.  As such, the theoretical and experimental communities are broadening their approach.  Light dark matter and axions, as well as axion like particles (ALPs) and dark photons, have recently received a lot of attention.  These particles are typically lighter than $\sim 1\,\text{GeV}$ and are very weakly coupled to the SM particles, which could make them good models to consider in an exploding black hole search.  However, these models typically only introduce a few degrees of freedom, so many photons would need to be observed to probe them.  In ref.~\cite{Baker:2021btk}, for example, around $10^5$ photons would need to be observed in HAWC to detect the presence of a light Dirac fermion dark matter candidate.  Since the black hole would need to evaporate very close to the Earth for this many photons to be detected (closer than $10^{-3}\,\text{pc}$), we would need to be very lucky to see one.
    
    \item Neutrino Masses -- While there are a range of models which explain the neutrino masses, perhaps the simplest explanation of the neutrino masses is the type-I see-saw, which introduces right-handed neutrinos at a high scale (typically around $10^{15}\,\text{GeV}$).  Since this scale is so high, it is very hard to probe and will only impact an exploding black hole in the very last instants of its life.  While there is interest in lower scale mechanisms, these typically only introduce a few new particles and the spectrum is very model dependent.
    
    \item Matter-Antimatter Asymmetry -- The main classes of models which aim to explain this asymmetry are electroweak baryogenesis and leptogenesis.  While electroweak baryogenesis requires a strongly first-order electroweak phase transition, which implies new degrees of freedom around the weak scale, relatively few new degrees of freedom are typically introduced.  Leptogenesis models also introduce relatively few new degrees of freedom and typically around $10^{15}\,\text{GeV}$ to tie in to seesaw explanations of the neutrino masses.
    
    \item Inflation -- While models of inflation are well motivated by cosmological data, they again typically only introduce a few new degrees of freedom.
    
    \item Models Motivated by Experimental Anomalies -- While a wide range of models have been proposed to explain existing experimental anomalies such as the $B$-physics anomalies and the anomalous magnetic moment of the muon, typically only one or two new particles are strictly required to explain the anomalies.  While some approaches also introduce a wide range of new particles, it is difficult to sensibly define `representative' models of this class given the wide range of approaches considered.
    
\end{itemize}

We now return to the models we consider to be of particular interest in the event of an observation of an exploding black hole.  We will summarise the particle content of these classes of models and define exemplary benchmarks. 

%-----------------------------------------------------------------------------
\subsection{Supersymmetry}
%-----------------------------------------------------------------------------

In supersymmetry (SUSY), at least one new copy of the SM particles, with fermions interchanged with bosons, is introduced, along with a second Higgs doublet.  Since no supersymmetric particles have yet been discovered, they must be somewhat heavier than the SM particles, which can be achieved through a variety of soft SUSY breaking mechanisms.  While the mass scale of the supersymmetric particles could in principle be as high as the Planck scale, there has been significant interest in low-scale supersymmetric models since as well as potentially providing a resolution to the gauge hierarchy problem, they can also improve gauge coupling unification and provide a thermal dark matter candidate.  Due to these theoretical successes, the prediction of new coloured states at the energy scale probed by the LHC, and the fact that sectors of the model can resemble a wide range of BSM theories, supersymmetry has been very widely studied.  Recent LHC searches mean that new uncoloured particles are likely all heavier than a few hundred GeV while coloured particles are likely all heavier than a TeV~\cite{ATLAS:2022rcw}.   Maintaining a solution to the hierarchy problem while passing these collider bounds leads to a little hierarchy problem, which can be somewhat ameliorated via, e.g., neutral naturalness~\cite{Batell:2022pzc} (where the top partner is uncoloured, so can be lighter than a TeV).

The Minimal Supersymmetric Standard Model (MSSM) introduces over 100 new degrees of freedom, along with around 100 new fundamental parameters.  There are several approaches taken to tackle this large parameter space: fixing relations between couplings at a high scale (such as the cMSSM and mSUGRA); choosing benchmark planes, lines and points (in high-scale or low-scale parameters); and using simplified topologies, where minimal assumptions on SUSY productions and decays are used to study certain new particles while other particles are taken to be heavy or weakly coupled.  While much of the MSSM parameter space remains viable, we here choose an illustrative benchmark point with all new particles at a similar mass scale.  Since we do not expect a dramatic difference between the signatures from an evaporating black hole when the particle masses are altered by a factor of a few, we focus on one benchmark point here and leave a more detailed study of a wider range of parameters for future work.  Beyond the MSSM, models with extra superfields are studied, such as the NMSSM (which aims to address the $\mu$-problem) and the $\nu$MSSM (which addresses neutrino masses).  However, since these typically only introduce a few extra degrees of freedom, they are expected to alter the signature of an evaporating black hole in a similar way to the MSSM.  Models like Split SUSY~\cite{Wells:2003tf,Arkani-Hamed:2004ymt,Giudice:2004tc,Arkani-Hamed:2004zhs} have widely separated mass scales, but then do not address the gauge hierarchy problem.  As such these models are less theoretically motivated than weak scale SUSY models.

As a benchmark we choose the $M_h^{125}$ benchmark of ref.~\cite{Bagnaschi:2018ofa} with $M_A = 2\,\text{TeV}$ and $\tan\beta = 20$ (see fig 1.~of~\cite{Bagnaschi:2021jaj}).  For this benchmark the squarks and the gluino (which are coloured so have a large LHC production cross-section) are safely above the current LHC bounds and all superpartners are chosen to be heavy so production and decay of the $125\,\text{GeV}$ Higgs boson is only mildly affected.  We use \texttt{SPheno}~\cite{Porod:2003um,Porod:2011nf} to generate the spectrum and we provide an overview of the new particles introduced in their mass basis in \cref{tab:susy-model}, noting which particles lead to additional secondary photons. In the SM it is not yet known whether the neutrinos are Dirac or Majorana fermions.  To be concrete, we choose them to be Majorana so that the SM contains 118 degrees of freedom. Since this can result from a type I see-saw with right-handed neutrinos heavier than $10^7\,\text{GeV}$, this choice does not necessarily significantly impact the observed signature of a black hole explosion.  We then use \texttt{pythia8} \cite{Sjostrand:2014zea} to compute their showering and hadronisation at various energies (detailed below), and make the secondary spectra as a function of black hole mass available at \href{https://github.com/PhysicsBaker/SecondarySpectra}{github.com/physicsbaker/secondaryspectra}.  We do not include the graviton or gravitinos because, as we will see in \cref{sec:particle-emission}, their emission is suppressed.

%-----------------------------------------------------------------------------
\begin{table}
\renewcommand{\arraystretch}{2}
\begin{tabular*}{\textwidth}{c @{\extracolsep{\fill}} cccc}
\hline
Particle & Spin & \# dof & Mass & Secondary Photons \\
\hline
\hline
SM & 0, 1/2, 1 & 118 & $M_\text{SM}$ & \checkmark\\
\hline
$\widetilde{g}$ & 1/2 & 16 & $\sim 2.5\,\text{TeV}$ & \checkmark\\ $\widetilde{\chi}^0_2$, $\widetilde{\chi}^0_3$, $\widetilde{\chi}^0_4$, $\widetilde{\chi}^\pm_1$, $\widetilde{\chi}^\pm_2$ & 1/2 & 14 & $\sim 1\,\text{TeV}$ & \checkmark\\
$\widetilde{\chi}^0_1$ & 1/2 & 2 & $\sim 1\,\text{TeV}$ & $\times$\\
$H^0$, $A^0$, $H^\pm$ & 0 & 4 & $\sim2\,\text{TeV}$ & \checkmark\\
\parbox{0.3\textwidth}{$\widetilde{d}_{L,R}$, $\widetilde{u}_{L,R}$,
$\widetilde{s}_{L,R}$, $\widetilde{c}_{L,R}$,\\
$\widetilde{e}_{L,R}$, $\widetilde{\mu}_{L,R}$,
$\widetilde{\tau}_{1,2}$,$\widetilde{\nu}_{eL}$, $\widetilde{\nu}_{\mu L}$, $\widetilde{\nu}_{\tau L}$}$\biggl\}$ & 0 & 66 & $\sim 2\,\text{TeV}$ & \checkmark
\\
$\widetilde{b}_{1,2}$, $\widetilde{t}_{1,2}$ & 0 & 24 & $\sim 1.5\,\text{TeV}$ & \checkmark\\
\hline
\end{tabular*}
\caption{Particle content of the benchmark MSSM Model with masses less than $\sim 10^7\,\text{GeV}$ and spin less than $3/2$, see text.}
\label{tab:susy-model}
\end{table}
%-----------------------------------------------------------------------------

%-----------------------------------------------------------------------------
\subsection{Large $N$ Species Solutions to the Hierarchy Problem}
%-----------------------------------------------------------------------------

Large $N$ species solutions to the hierarchy problem were developed in refs.~\cite{Dvali:2007hz,Dvali:2007wp,Calmet:2008tn,Dvali:2009ne} as an alternative to supersymmetric, composite Higgs and extra-dimensional models. They introduce $N \sim 10^{32}$ species beyond the SM in order to reduce the scale of gravity.  The apparent Planck scale, $M_\text{Pl}$, is related to the true scale where gravity becomes strongly coupled, $M_\ast$, via the relation $M_\text{Pl}^2 \gtrsim N M_\ast^2$. The new species are at or below the weak scale and couple only gravitationally. In this scenario, black holes can form at the TeV scale and could be produced and probed at the LHC. The signature of an exploding black hole in models with $M_\ast \ll M_\text{Pl}$ would be strikingly different to models with $M_\ast$ above experimental sensitivity. Evaporation would stop at $M_\ast$ and could allow us to infer the true scale of gravity. We leave a detailed discussion of this scenario to future work. 

%-----------------------------------------------------------------------------
\begin{table}
\renewcommand{\arraystretch}{2}
\begin{tabular*}{0.75\textwidth}{c @{\extracolsep{\fill}} ccccc}
\hline
Particle    & Spin      & \# dof & Mass & Secondary Photons \\
\hline
\hline
SM          & 0, 1/2, 1   & 118              & $M_\text{SM}$                & \checkmark\\
\hline
Reheaton    & 0 or 1/2  & 1 or 2                        & weak scale      & \checkmark\\
Higgs       & 0     & $5 \times 10^{15}$   & $\Lambda_{\text{QCD}}$        & $\times$\\
Photons     & 1     & $2 \times 10^{16}$   & 0                             & $\times$\\
Gluons      & 1     & $16\times 10^{16}$   & ${}^\ast\Lambda_{\text{QCD}}$                             & $\times$\\
$W^\pm$/$Z$         & 1     & $9 \times 5 \times 10^{15}$   & $\Lambda_{\text{QCD}}$            & $\times$\\
Neutrinos   & 1/2   & $3 \times 2\times 5 \times 10^{15}$    & $<100\,$eV                     & $\times$\\
Charged leptons     & 1/2   & $3\times 4 \times 5 \times 10^{15}$    & $<100\,$eV            & $\times$\\
Quarks      & 1/2   & $6\times 12 \times 5 \times 10^{15}$   & ${}^\ast\Lambda_{\text{QCD}}$            & $\times$\\
\hline
\end{tabular*}
\caption{Particle content of the $N$naturalness benchmark model (with $N=10^{16}$) with masses less than the weak scale.  We omit SM copies with a vev above the weak scale, see text.  ${}^\ast\Lambda_{\text{QCD}}$ denotes coloured particles with a mass below $\Lambda_{\text{QCD}}$ which we assume are only emitted above $\Lambda_{\text{QCD}}$, see text in  \cref{sec:particle-emission}.
Assuming $N$naturalness, black holes evaporating today would have a temperature around $100\,\text{eV}$ shortly after the Big Bang, so we treat particles with masses below this scale as massless.
}
\label{tab:$N$naturalness-model}
\end{table}
%-----------------------------------------------------------------------------

Here we focus on $N$naturalness \cite{Arkani-Hamed:2016rle}, a model which reduces $N$ (and raises $M_\ast$) while simultaneously solving the gauge hierarchy problem. $N$naturalness introduces $N-1$ new sectors with the same particle content as the SM. In half of the SM copies electroweak symmetry is broken at the QCD scale, while the other half contains particles which are heavier than the weak scale. In the simplest scenario, the Higgs mass parameter in all $N$ sectors is given by
\begin{equation}
(m_H^2)_i = (2i+1) (m_H^2)_\text{SM} , \qquad -\frac{N}{2} < i < \frac{N}{2} \,,
\end{equation}
where $(m_H^2)_{\text{SM}} = -(88\,\text{GeV})^2$ (so $i=0$ corresponds to the SM particles). Sectors with $(m_H^2)_i < 0$ (for $i>0$) are similar to the SM, but the Higgs vacuum expectation value (vev) and all associated particle masses are rescaled by $\sqrt{(2i+1)}$ (or $\sim i$ in the case of Majorana neutrinos). In sectors with $(m_H^2)_i > 0$ (for $i<0$) the electroweak symmetry is broken by QCD condensates and all fermionic and gauge degrees of freedom have masses $\lesssim \Lambda_\text{QCD}$. 

In this scenario the hierarchy problem is solved with cosmological dynamics which ensure that only (or almost only) the $i=0$ copy of the SM is populated by reheating. A reheaton is introduced which, in the simplest models, is taken to be a real scalar or a Majorana fermion and couples to two Higgses or with a Yukawa-like coupling to each of the $i$ sectors. The reheaton has a branching ratio of $\text{BR}_i \sim (m_H)^{-\alpha}_i$ with $\alpha > 0$ which ensures that most of the reheaton's energy is deposited into the SM sector, i.e., the sector with $i=0$. A black hole would, however, radiate particles from all $i$ sectors via Hawking radiation. 

For our benchmark model we require the black hole to be SM-like up to a scale of $10^7\,\text{GeV}$.  Taking $M_\ast \gtrsim 10^7\,\text{GeV}$ implies that $N \lesssim 10^{22}$.  We also require that no further unknown, unspecified new physics exists below $10^7\,\text{GeV}$, since we need the model to be calculable up to this scale. This leads to $N\gtrsim 10^{10}$. Both conditions are satisfied for the benchmark we choose, $N=10^{16}$, where $M_\ast \sim 10^{10}\,\text{GeV}$. This model has $5\times 10^{15}$ SM copies with masses $\lesssim \Lambda_{\text{QCD}}$ but only $5\times 10^{9}$ SM copies with masses between the SM masses and $10^7\,\text{GeV}$. We can therefore neglect the massive copies.  The particle spectrum of the light sectors is summarised in \cref{tab:$N$naturalness-model}.

To simplify our analysis, we further neglect the effect of the reheaton. The one or two degrees of freedom of the reheaton will be negligible compared to the $N$ SM copies, and it gives a subdominant contribution to the secondary spectra (compared to the SM particles).  This assumption decouples the $N$ sectors from each other and from the SM, so only the SM particles produce a significant flux of secondary photons. This assumptions also avoids introducing any model dependence, through the specific mass dependent couplings of the reheaton.

%-----------------------------------------------------------------------------
\subsection{String Inspired Models}
%-----------------------------------------------------------------------------

String theory is only self-consistent in higher dimensional space-times.\footnote{In contrast to large extra-dimensional models, the extra dimensions are very small, so black hole dynamics significantly below the Planck scale are not dramatically affected.}  
Compactification of the underlying space-time manifold typically leads to many massless scalars (moduli).  These are typically made massive via a variety of mechanisms to evade cosmological bounds and fifth-force constraints (this process is known as moduli stabilisation).  The two main classes of moduli stabilisation techniques are KKLT~\cite{Kachru:2003aw} and the Large Volume Scenario (LVS)~\cite{Balasubramanian:2005zx}.  KKLT typically results in moduli heavier than $10^6$ GeV, so we will here focus on the LVS.  

%-----------------------------------------------------------------------------
\begin{table}
\renewcommand{\arraystretch}{2}
\begin{tabular*}{0.75\textwidth}{c @{\extracolsep{\fill}} cccc}
\hline
Particle & Spin & \# dof & Mass & Secondary Photons \\
\hline
\hline
SM & 0, 1/2, 1 & 118 & $M_\text{SM}$ & \checkmark\\
\hline
$a_b$ & 0 & 3 & $\ll \text{GeV}$ & $\times$\\
$\tau_b$ & 0 & 3 & $M_\text{Pl}/\mathcal{V}^{3/2}$ & $\times$\\
$\widetilde{G}$, $T$, $U$, $S$ & $0,\,3/2$ & $\sim 400$ & $M_\text{Pl}/\mathcal{V}$ & $\times$\\
\hline
\end{tabular*}
\caption{Particle content of the string theory inspired models with masses less than $\sim 10^7\,\text{GeV}$.  The volume $\mathcal{V}$ is a dimensionless number which we take to be in the range $10^{12} - 10^{15}$.}
\label{tab:string-model}
\end{table}
%-----------------------------------------------------------------------------

The masses of the lightest moduli scale with the volume of the internal manifold, $\mathcal{V}$, which is dimensionless when expressed in units of the string length, as shown in \cref{tab:string-model}~\cite{Cicoli:2012aq}.  While volumes in the range $10^5 \lesssim \mathcal{V} \lesssim 10^{15}$ are typically considered, we will focus on scenarios with  $\mathcal{V} \gtrsim 10^{12}$ so that new degrees of freedom appear below $10^6\,\text{GeV}$.  As benchmarks we will choose $\mathcal{V} = 10^{15}$ and $\mathcal{V} = 10^{13}$.  The lightest modulus is the volume axion, $a_b$, which typically has a mass much lower than the GeV scale and which we will treat as massless.  The next lightest is the volume modulus, $\tau_b$, which has a mass around $10^{-4}\,(10^{-1})\,\text{GeV}$ for  $\mathcal{V} = 10^{15}\,(10^{13})$.  The K\"ahler moduli, $T$, the complex structure moduli, $U$ and the axiodilaton, $S$, all have similar masses to the gravitino, $\widetilde G$, which is around $2\times 10^3\,(2\times 10^5) \,\text{GeV}$ for  $\mathcal{V} = 10^{15}\,(10^{13})$.  Since there are significantly more degrees of freedom in the moduli than in the gravitino, we neglect the gravitino contribution.

String theory may also give rise to hundreds or thousands of sub-eV axions~\cite{Arvanitaki:2009fg,Cvetic:2022fnv}.  While the details are quite model dependent, these would have a significant impact on black hole evaporation and would be a prime candidate for future work.

%-----------------------------------------------------------------------------
\subsection{Dark Sectors}
%-----------------------------------------------------------------------------

In ref.~\cite{Baker:2021btk} we studied dark sector models motivated by Mirror Dark Matter~\cite{Foot:1991bp}.  In that work we introduced $N$ copies of the SM degrees of freedom, but all with a common mass scale $\Lambda_\text{DS}$.  We label these models DS$(N,\Lambda_\text{DS})$.  The particle content of these models is given in \cref{tab:dark-sector-model}.  The new particles are taken to only interact weakly with the SM particles, so that no (or an insignificant number of) secondary photons are produced.  In this work we study three concrete benchmarks: DS$(10,10^2\,\text{GeV})$, DS$(2,2\times10^3\,\text{GeV})$ and DS$(1,10^4\,\text{GeV})$.

%-----------------------------------------------------------------------------
\begin{table}
\renewcommand{\arraystretch}{2}
\begin{tabular*}{0.75\textwidth}{c @{\extracolsep{\fill}} cccc}
\hline
Particle & Spin & \# dof & Mass & Secondary Photons \\
\hline
\hline
SM & 0, 1/2, 1 & 118 & $M_\text{SM}$ & \checkmark\\
\hline
DS & 0, 1/2, 1 & $118\, N$ & $\Lambda_\text{DS}$ & $\times$\\
\hline
\end{tabular*}
\caption{Particle content of the Dark Sector Models.}
\label{tab:dark-sector-model}
\end{table}
%-----------------------------------------------------------------------------

%=============================================================================
\section{Particle Emission from an Evaporating Black Hole}
\label{sec:particle-emission}
%=============================================================================

Black holes radiate particles of type $i$ with energy $E$ via Hawking radiation at the rate~\cite{Hawking:1974rv,Hawking:1974sw}
\begin{align}
    \label{eq:d2NdEdt}
    \frac{d^2 N^i_\text{p}}{dtdE} =&\, \frac{n^i_\text{dof} \, \Gamma^i}{2\pi(e^{E/T}\pm 1)} \,,
\end{align}
where $n^i_\text{dof}$ is the number of degrees of freedom of particle $i$, $+$ $(-)$ corresponds to fermions (bosons), $\Gamma^i$ is a greybody factor, $T =1/(8\pi \, G M)$ is the temperature of the black hole in units where $\hbar=c=\kappa_\text{B}=1$, $G$ is the gravitational constant and $M$ is the black hole mass. For the SM particles, $n_q = 12$ for each quark, $n_\ell = 4$ for each charged lepton, $n_\nu = 2$ for each neutrino (assuming Majorana neutrinos), $n_g = 16$ for the gluon, $n_\gamma = 2$ for the photon, $n_Z = 3$ for the $Z$ boson, $n_{W^\pm} = 6$ for the $W$ boson and $n_h = 1$ for the Higgs boson.  In general the greybody factors depend on the emitted particle's energy, mass and spin and the black hole's mass, spin and charge.  However, black holes emit angular momentum and charge faster than mass~\cite{Page:1976ki,Zaumen:1974,Carter:1974yx,Gibbons:1975kk,Page:1976df,Page:1977um} so near the end of their life they will be approximately Schwarzschild (zero angular momentum and no charge).  While the emission in general depends on the mass of the emitted particle, emission of particles with $E < m$ is heavily suppressed at significant distances from the black hole~\cite{Jannes:2011qp} and any mass corrections when $E\sim m$ only affect very few particles, so we set the greybody factor to zero when $E < m$ and use the massless expression for $E > m$. Then, the greybody factor only depends on the ratio of $E$ to the black hole's mass, with a threshold at the particle's mass. We take the greybody factors from the publicly available \texttt{BlackHawk} code~\cite{Arbey:2019mbc,Arbey:2021mbl}.  The exponential factor in the emission rate, \cref{eq:d2NdEdt}, ensures that the emission rate for particles with $E \gg T$ is heavily suppressed.  It is important to note that the emission rate is independent of any non-gravitational interactions.  As such, all particles present in nature will be emitted according to \cref{eq:d2NdEdt}, even if they are only coupled gravitationally to the SM particles.

%-----------------------------------------------------------------
\begin{figure}
    \centering
    \includegraphics[width=0.6\textwidth]{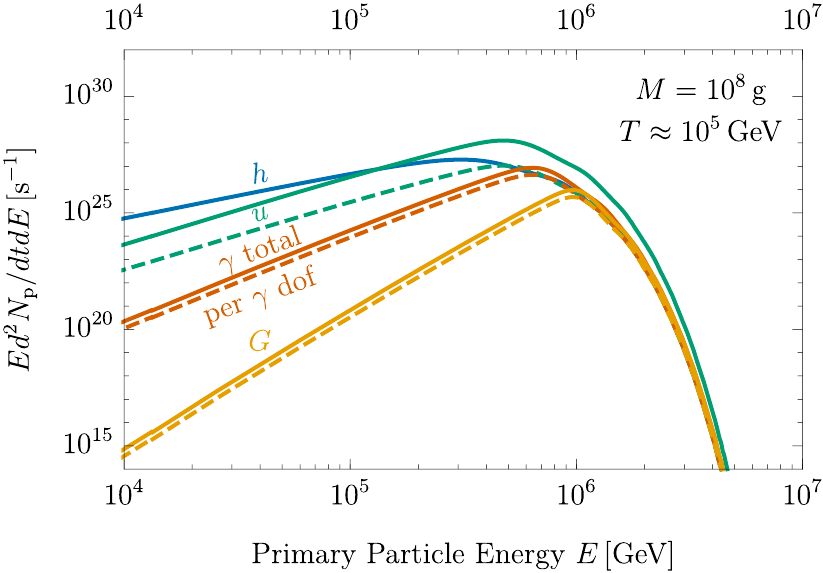}
    \caption{Representative primary spectra for particles of spin 0 (the Higgs boson, $h$), spin $1/2$ (the up quark, $u$), spin 1 (the photon, $\gamma$) and spin 2 (the graviton, $G$) for a black hole of mass $10^8\,\text{g}$.  The dashed lines show the spectra per degree of freedom while the solid lines include the factors 
    $n_\text{dof}^h = 1$, $n_\text{dof}^u = 12$, $n_\text{dof}^\gamma = 2$ and $n_G = 2$, respectively. 
    }
    \label{fig:primary}
\end{figure}
%-----------------------------------------------------------------

In \cref{fig:primary} we show the primary spectra for Higgs bosons (spin 0), up quarks (spin 1/2) and photons (spin 1) for a black hole with a mass of $10^8\,\text{g}$.\footnote{Since we plot on a log scale in energy, we in fact plot the spectra multiplied by the energy so that the number of photons at different energies can be easily compared by eye.}  The solid lines show the total rate for that species when the number of degrees of freedom are taken into account ($n_\text{dof}^h = 1$, $n_\text{dof}^u = 12 = 2 \,\text{(spin)} \times 2 \,\text{(particle/anti-particle)} \times 3 \,\text{(colour)}$, $n_\text{dof}^\gamma = 2$), while the dashed lines show the rate per degree of freedom.  We see that particles of higher spin have a smaller emission rate at energies $E \lesssim T$.  The suppression is greater for particles of higher spin, and lead to the spectra peaking at higher energies for higher spin particles.  Since the graviton rate is so suppressed, we omit it from our analysis.  Similarly, in the MSSM case we omit the gravitino.

%-----------------------------------------------------------------
\begin{figure}
    \centering
    \begin{tabular}{cc}
    \includegraphics[width=0.475\textwidth]{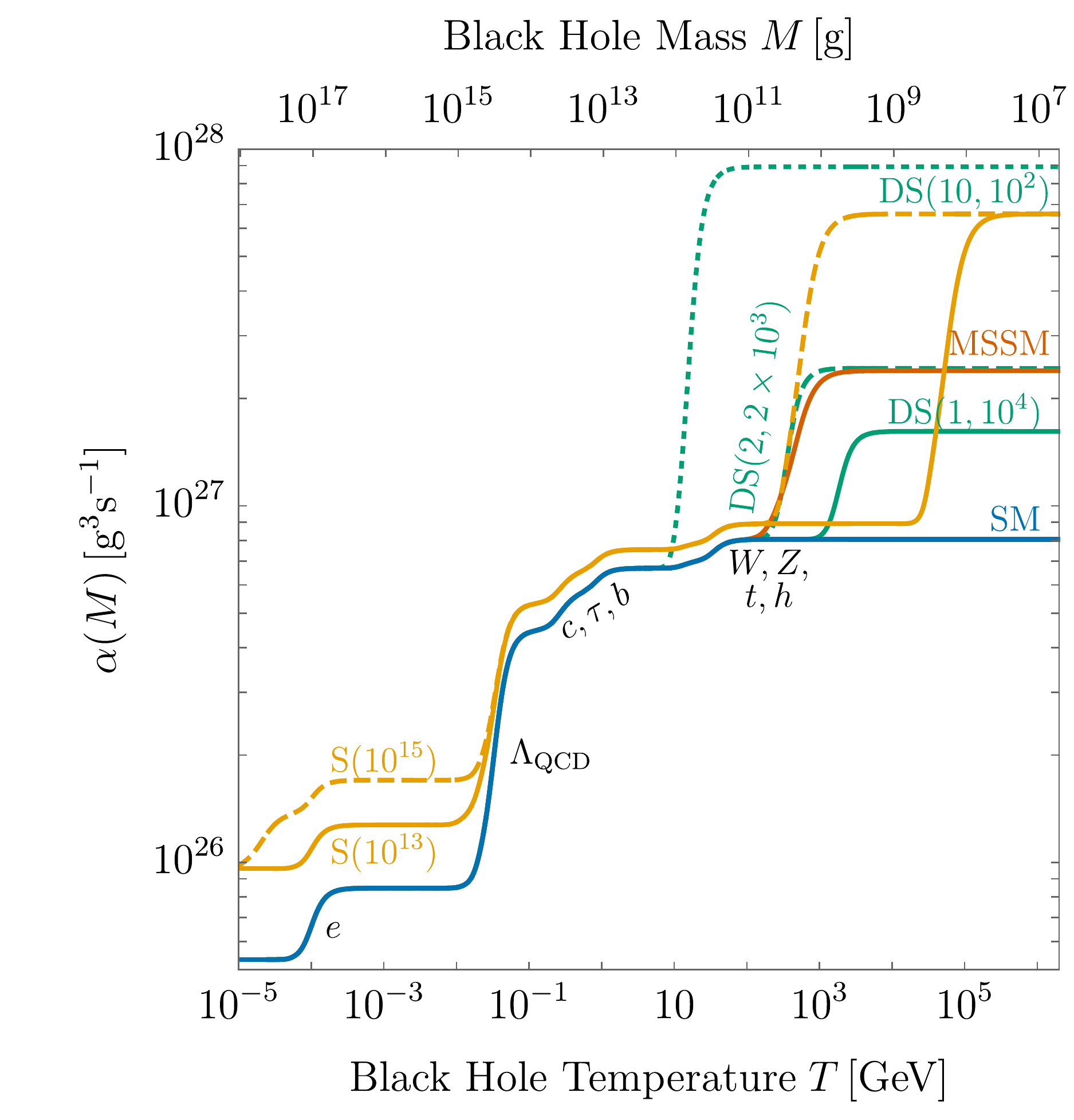}&
    \includegraphics[width=0.475\textwidth]{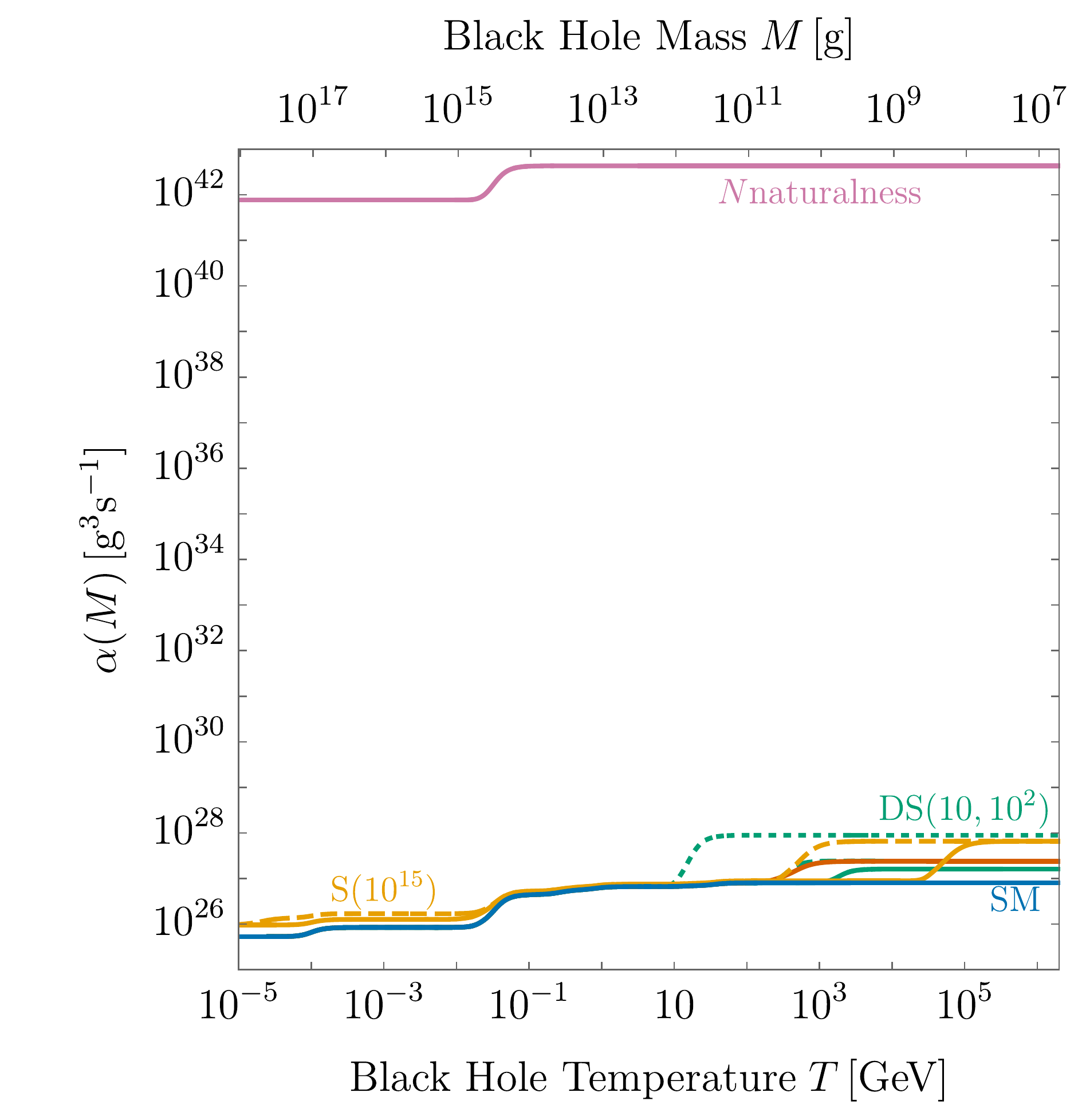}
    \end{tabular}
    \caption{The function $\alpha(M)$ which accounts for all directly emitted particle species, for our benchmark models excluding (left) and including (right) $N$naturalness.
    }
    \label{fig:alpha}
\end{figure}
%-----------------------------------------------------------------

Neglecting accretion due to gas, light, dark matter, etc, energy conservation implies that as particles are emitted from the black hole, its mass evolves according to~\cite{Page:1976df}
\begin{align}
    \label{eq:dMdt}
    \frac{dM}{dt} =&\, -\frac{\alpha(M)}{M^2} \,,
\end{align}
where
\begin{align}
    \alpha(M) =&\, M^2 \sum_i \int_{m_i}^\infty\frac{d^2 N_\text{p}}{dtdE}(M,E) \,E \, dE\,,
\end{align}
and the sum is over all particles present in the theory.  We explicitly show that we take the integral to be over $m_i < E$ as a reminder that we set the greybody factor to zero when $m_i > E$.  The function $\alpha(M)$ is shown in \cref{fig:alpha} for the benchmark models we consider.  We see that $\alpha(M)$ is constant away from particle thresholds, but increases in steps when the temperature of the black hole becomes higher than the mass of the particles.  The curve for the SM is the smallest that $\alpha(M)$ can be, since the SM contains only those particles that are known to exist (for minimality we consider the SM neutrinos to be Majorana states).  We see that the string theory inspired models have a larger $\alpha(M)$ than the SM at temperatures lower than the QCD scale, due to the light volume axion, $\theta_b$, and the volume modulus, $\tau_b$, which has a mass around $10^{-4}\,\text{GeV}$ when $\mathcal{V}=10^{15}$.  Around and below $0.1\,\text{GeV}$ QCD is strongly coupled, so perturbation theory cannot be used to calculate QCD effects.  Furthermore, there is debate in the literature as to whether black holes can emit free quarks below the QCD scale or only above the pion mass scale, the lightest colourless QCD bound state~\cite{MacGibbon:1990zk,MacGibbon:1991tj,Coogan:2020tuf,Arbey:2021mbl}.  We assume that only colourless states such as the pion can be emitted, and use an effective mass of $m_i = \Lambda_\text{QCD} \sim 0.2\,\text{GeV}$ for the gluon and the $u$, $d$ and $s$ quarks in the SM and for the light coloured states in our $N$naturalness benchmark model.

Above the QCD scale, the first model to show a step is the dark sector DS$(10,10^2\,\text{GeV})$ benchmark, where we assume a dark sector scale of $\Lambda_\text{DS} = 100\,\text{GeV}$ and 10 copies of the SM degrees of freedom.  The $\mathcal{V}=10^{15}$ string inspired model, the MSSM benchmark and DS$(2,2\times10^3\,\text{GeV})$ all feature a rise just below $T\sim 1\,$TeV, when particles of this mass begin to be emitted.  The step for the $\mathcal{V}=10^{15}$ string inspired model is larger than the MSSM and DS$(2,2\times10^3\,\text{GeV})$ benchmarks due to the greater number of new degrees of freedom.  The dark sector DS$(1,10^4\,\text{GeV})$ benchmark features a step between $10^3$ and $10^4\,\text{GeV}$ while the $\mathcal{V}=10^{13}$ string theory inspired model has the same number of new degrees of freedom as the $\mathcal{V}=10^{15}$ string inspired model, but which turn on at a higher temperature, around $10^5\,\text{GeV}$.  In \cref{fig:alpha} (right) we also show $\alpha(M)$ for the $N$naturalness benchmark.  We see that it is many orders of magnitude larger than the other benchmarks, due to the huge number of new degrees of freedom introduced.  It increases by an order of magnitude, from $8\times 10^{41}\,\text{g}^3 \text{s}^{-1}$ below the QCD scale to $4\times 10^{42}\,\text{g}^3 \text{s}^{-1}$ above the QCD scale (as the coloured particles, the electroweak gauge bosons and the Higgs of the copies where electroweak symmetry is broken at the QCD scale begin to be emitted).

%-----------------------------------------------------------------
\begin{figure}
    \centering
    \begin{tabular}{cc}
    \includegraphics[width=0.475\textwidth]{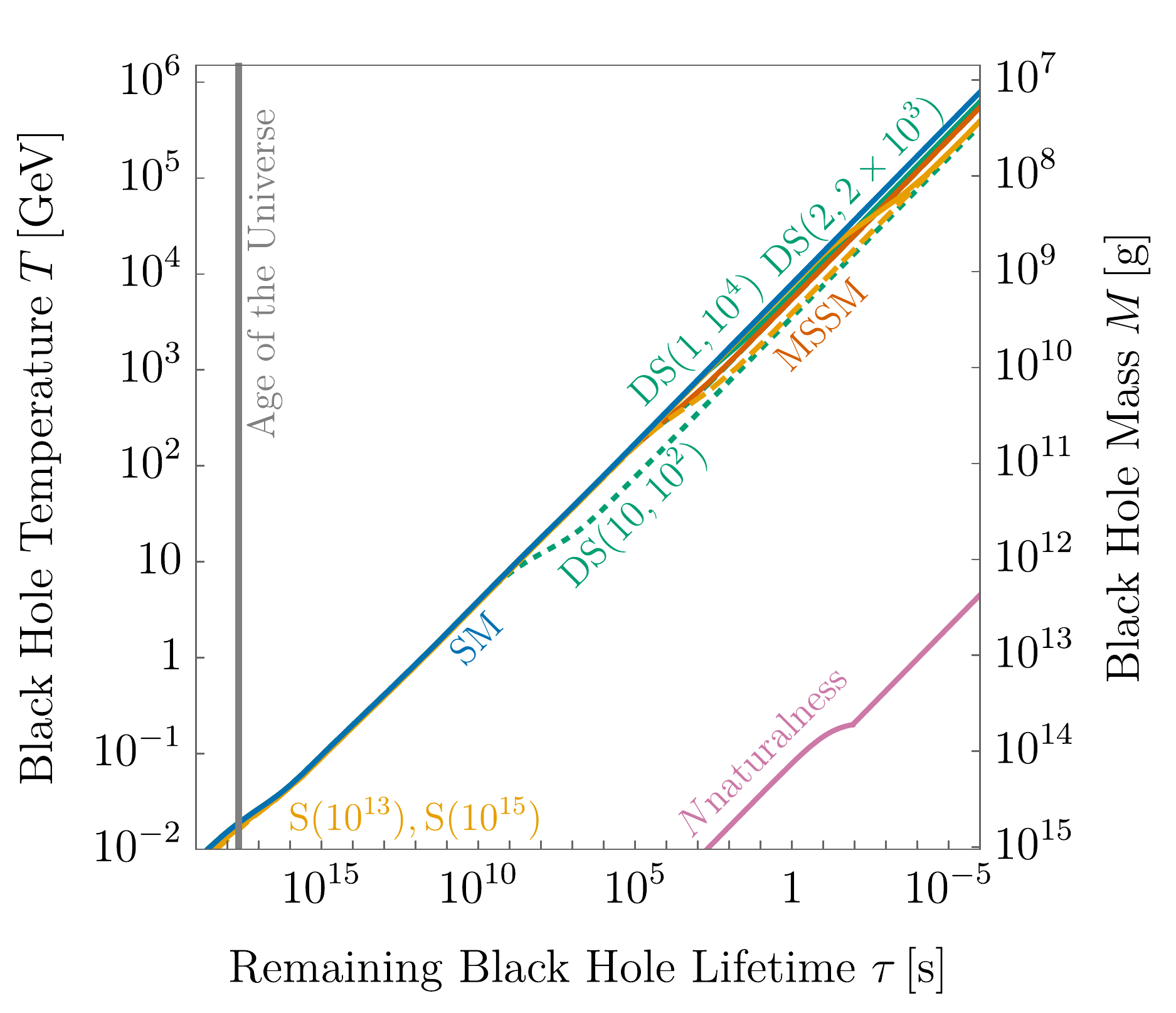}&
    \includegraphics[width=0.475\textwidth]{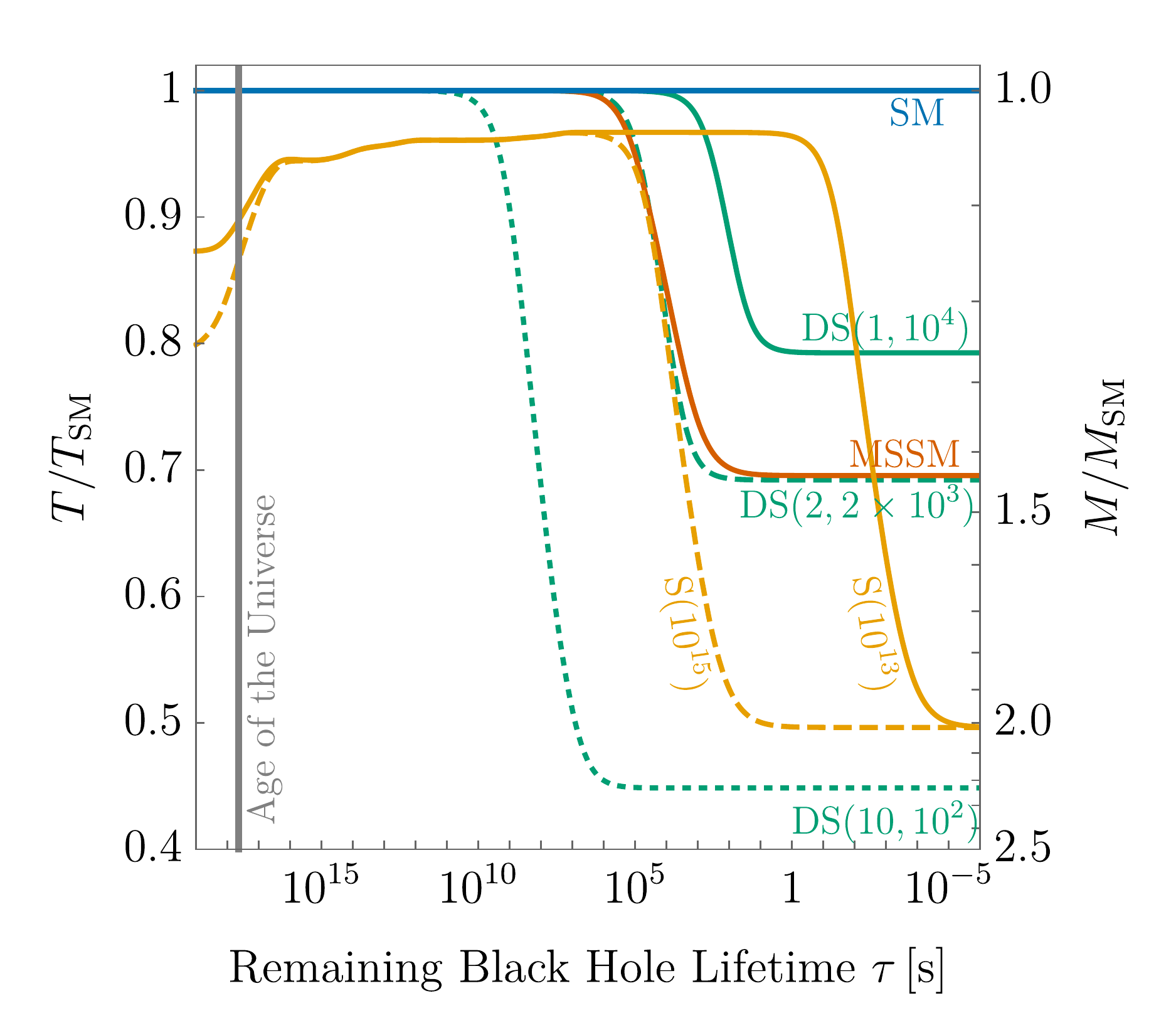}\\
    \includegraphics[width=0.475\textwidth]{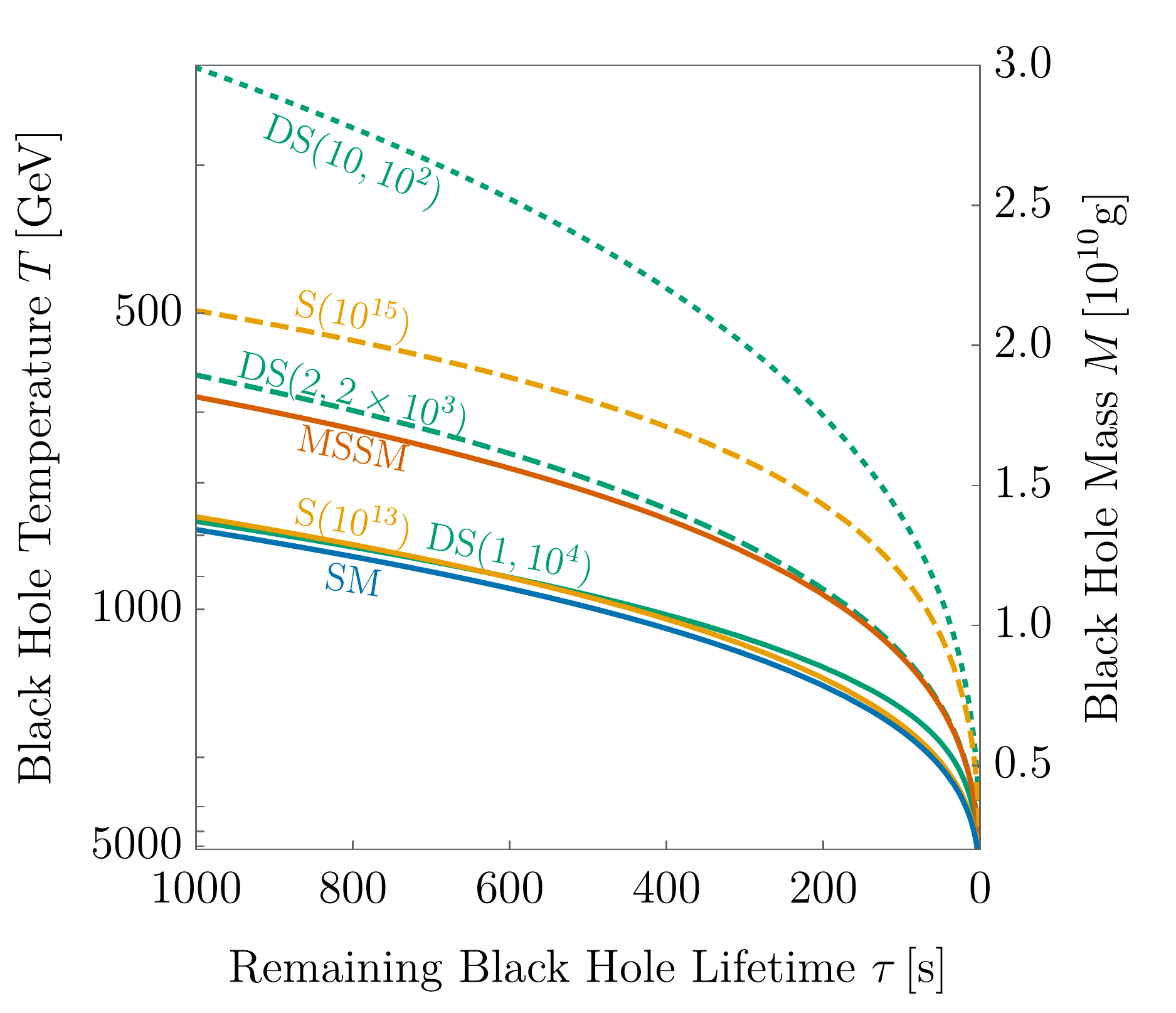}&
    \includegraphics[width=0.475\textwidth]{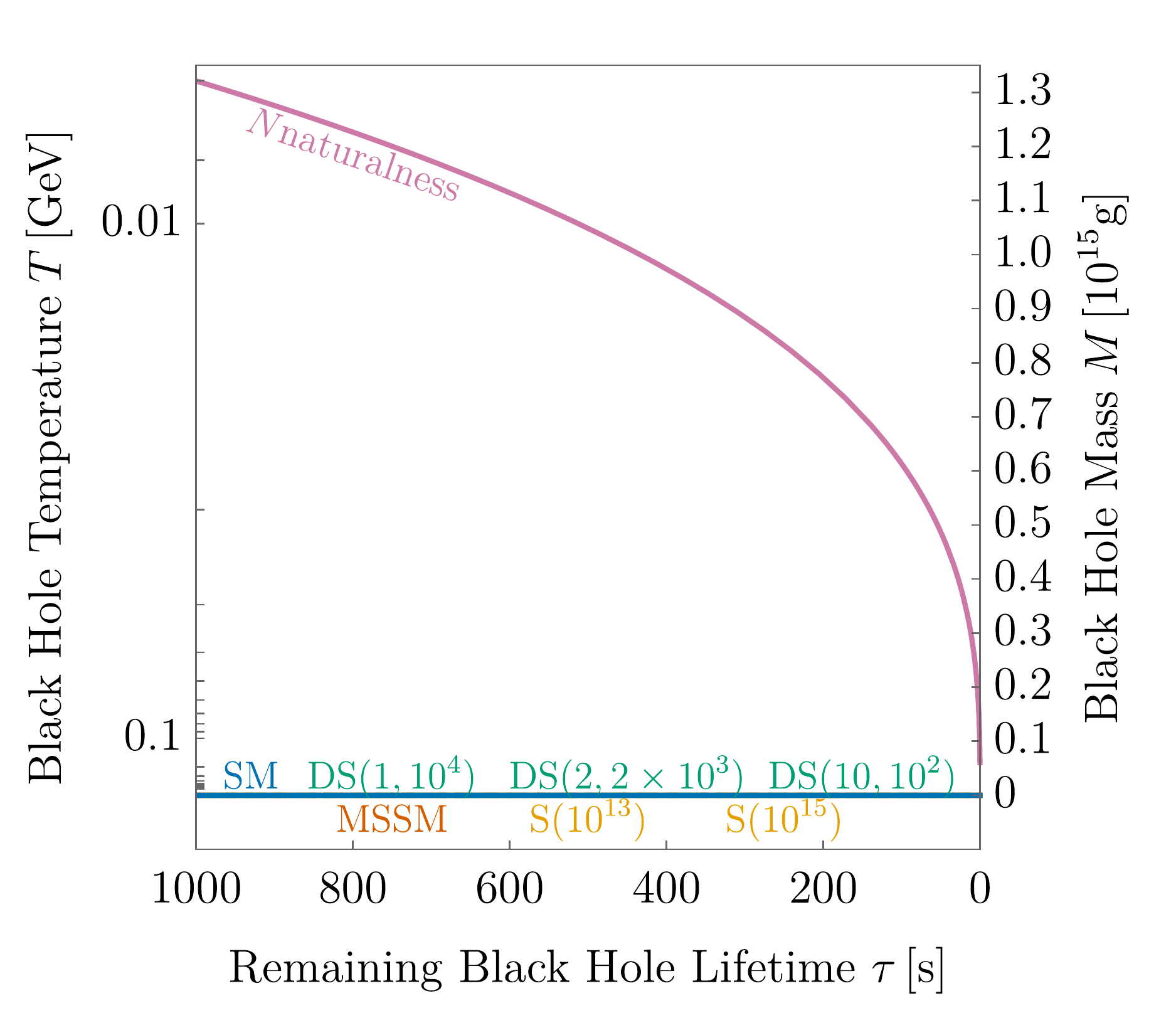}
    \end{tabular}
    \caption{Top left: the black hole temperature and mass as a function of the remaining black hole life time, $\tau$, for our benchmark models.  Top right: the ratio of the black hole temperature to the SM expectation at the same $\tau$ for our benchmark models.
    Bottom row: the back hole temperature and mass on a linear scale, excluding (bottom left) and including (bottom right) the $N$naturalness benchmark model.
    }
    \label{fig:m-tau}
\end{figure}
%-----------------------------------------------------------------

Once $\alpha(M)$ has been computed for a given theory, \cref{eq:dMdt} determines the evolution of the black hole temperature (or equivalently, its mass) with time (neglecting any accretion onto the black hole).  The result can be seen in \cref{fig:m-tau} for our benchmark scenarios, where time is parameterised in terms of the time remaining before the end point of the black hole explosion, $\tau$.  The top panels show the behaviour in log $\tau$, while the lower panels show it linear in $\tau$ for the final $10^3\,\text{s}$ of the black hole's life.  If $\alpha(M)$ is constant, then \cref{eq:dMdt} shows that $M \propto \tau^{1/3}$.  This is reflected in the general slope of the evolution seen in the top left panel.  We see that when $T \sim 10\,\text{GeV}$ and $M_\text{BH} \sim 10^{12}\,\text{g}$ the slope temporarily becomes less steep for the DS$(10,10^2\,\text{GeV})$ benchmark.  This is because this model has new degrees of freedom that start to be emitted at this temperature, as seen in \cref{fig:alpha}.  While it may seem counter-intuitive that the temperature is only lower at later times, this is a something of an artefact due to plotting on a log scale.  In the DS$(10,10^2\,\text{GeV})$ scenario the black hole is cooler and heavier than in the SM scenario at the same $\tau$, but the disparity only becomes appreciable when the difference in mass is of a similar size to the mass itself.  Similarly, other benchmark models have an increased rate of mass loss when $T \gtrsim 100\,\text{GeV}$.  We see that the large number of new degrees of freedom in the $N$naturalness benchmark leads to a much greater rate of mass loss, and as such a lower black hole temperature and a larger black hole mass over the majority of the lifetime of the black hole.

In the top right panel of \cref{fig:m-tau} we show the ratio of the black hole temperature at a given time to the temperature expected assuming the SM, at the same time.  This shows more clearly the temperature deviations from the SM expectation.  Shortly after the Big Bang, the string inspired benchmarks have a temperature that is $\sim 10 - 15\%$ lower, due to the new light degrees of freedom in these models.  The gap closes somewhat by $10^{16}\,\text{s}$ as the large number of degrees of freedom that become available around the QCD scale begin to be emitted, reducing the impact of the light string degrees of freedom.  After that, the benchmarks move to lower temperatures than the SM expectation when the temperature rises above the mass thresholds.  We see that once the thresholds are passed, the black hole temperature is $\sim 25 - 55\%$ cooler than would be expected under the SM scenario.  This is a significant deviation.

In the lower panels of \cref{fig:m-tau}, we see the temperature (and mass) evolution in the last 1000 seconds of a black hole's life.  Under the assumption of the SM, the black hole would have a temperature around $800\,\text{GeV}$ when $\tau = 1000\,\text{s}$.  The benchmarks which have significant numbers of new degrees of freedom at approximately this scale have a significantly higher rate of mass loss with time.  The DS$(1,10^4\,\text{GeV})$ and $\mathcal{V}=10^{13}$ string benchmarks only have many new degrees of freedom at scales $\gg 1\,\text{TeV}$, so the evolution is not markedly different from the SM scenario over the last $1000\,\text{s}$ (although they do diverge significantly on shorter time scales).  In the lower right panel we see that the temperature is around five orders of magnitude smaller in the $N$naturalness benchmark, as the rate of mass loss is many orders of magnitude larger.

Assuming only the particles present in the SM, a black hole with a mass of $5.6\times10^{14}\,\text{g}$ produced shortly after the Big Bang would be evaporating today.  This is not significantly changed in models which only introduce new degrees of freedom above the GeV scale (DS$(1,10^4\,\text{GeV})$, DS$(2,2\times 10^3\,\text{GeV})$, DS$(10,10^2\,\text{GeV})$, and the MSSM benchmark) since the black hole spends most of its life with a temperature around 100 MeV.   The string theory inspired benchmarks introduce three sub-GeV degrees of freedom, which lift the initial mass (assuming a primordial origin) to $6.5\times10^{14}\,\text{g}$ and $6.2\times10^{14}\,\text{g}$ for $\mathcal{V}=10^{15}$ and $10^{13}$, respectively.  The $N$naturalness model introduces a very large number of sub-GeV degrees of freedom.  As such, the evaporation of the black hole differs significantly: in this scenario, a black hole with a primordial mass of  $1.0\times10^{20}\,\text{g}$ would be evaporating today.

%=============================================================================
\section{The Primary and Secondary Photon Spectra}
\label{sec:photon-spectra}
%=============================================================================

Of all the particles emitted by an evaporating black hole, the most important one from an observational point of view is the photon.  These are emitted both as primary particles, as seen in \cref{fig:primary}, but also as secondary particles as other particles decay, shower and/or hadronise.  In this section we discuss the primary and secondary photon spectra in the various benchmark models.

\begin{figure}
    \centering
    \includegraphics[width=0.6\textwidth]{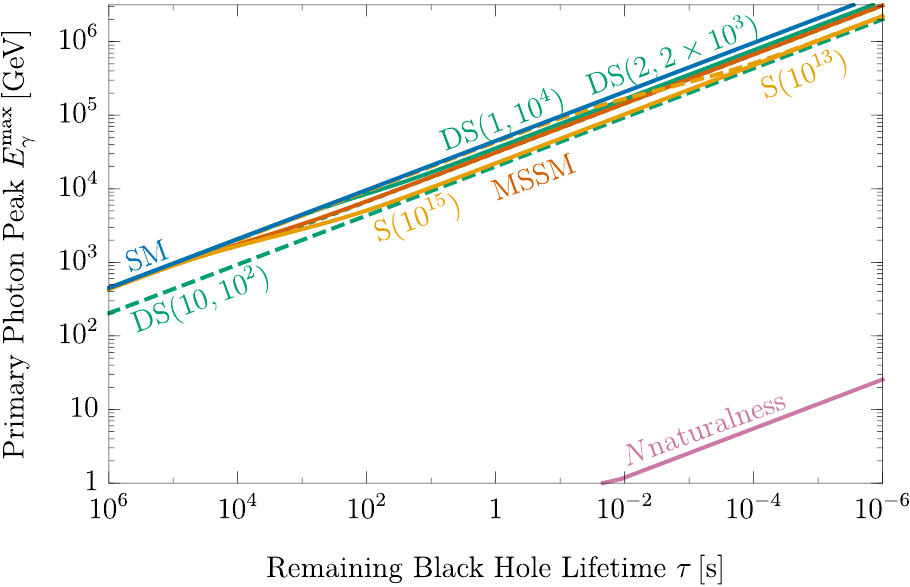}
    \caption{The peak energy of the primary photons as a function of the remaining black hole life time for our benchmark models.
    }
    \label{fig:Ep-tau}
\end{figure}

The primary photon spectra for a Schwarzschild black hole of a given mass is given by \cref{eq:d2NdEdt} with $n_\text{dof} = 2$ and where $\Gamma^\gamma$ is the greybody factor for a spin 1 particle, see \cref{fig:primary}.  Given the mass evolution computed in the previous section, we can then find the primary photon spectrum at any time.  In \cref{fig:Ep-tau} we show the position of the peak of the primary photon spectra as a function of $\tau$ for the benchmark models.  As there is a one-to-one correlation between the black hole temperature, mass and the position of the peak, the features are very similar to those seen in \cref{fig:m-tau}.  For models with more degrees of freedom, the black hole is significantly cooler at later times, and so emits photons of lower energies at later times.  The $N$naturalness benchmark only emits photons above $0.1\,\text{GeV}$ in the last 100 seconds.  We will return to this point in \cref{sec:sm-bsm}.

%-----------------------------------------------------------------
%-----------------------------------------------------------------

\begin{figure}
    \centering
    \begin{tabular}{cc}
    \includegraphics[width=0.475\textwidth]{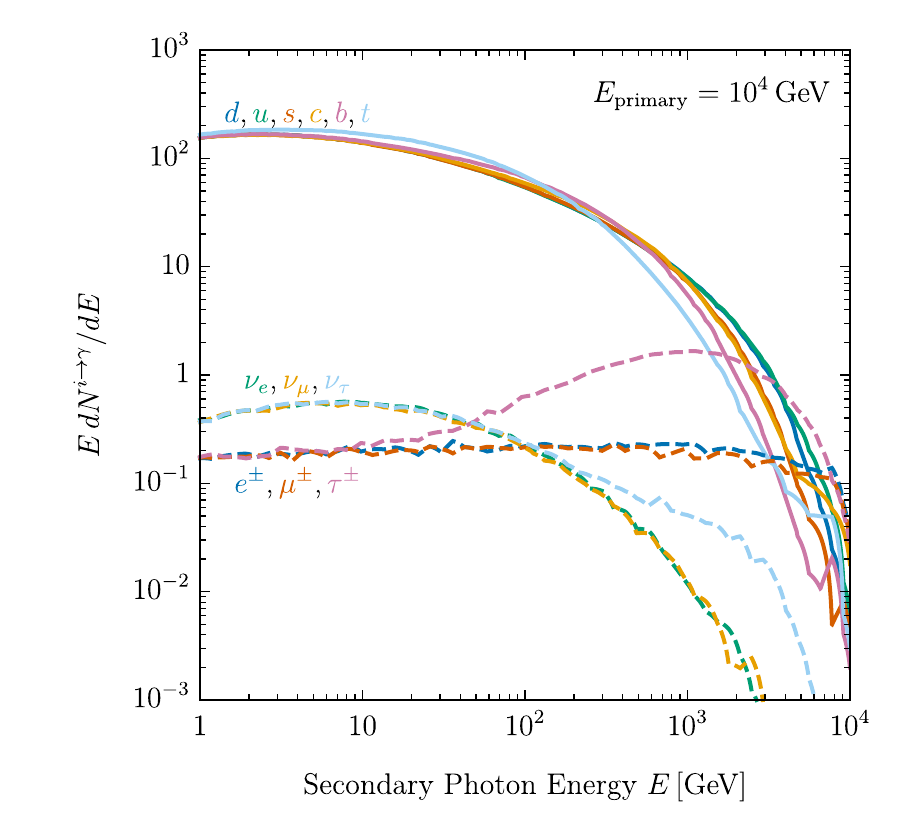}&
    \includegraphics[width=0.475\textwidth]{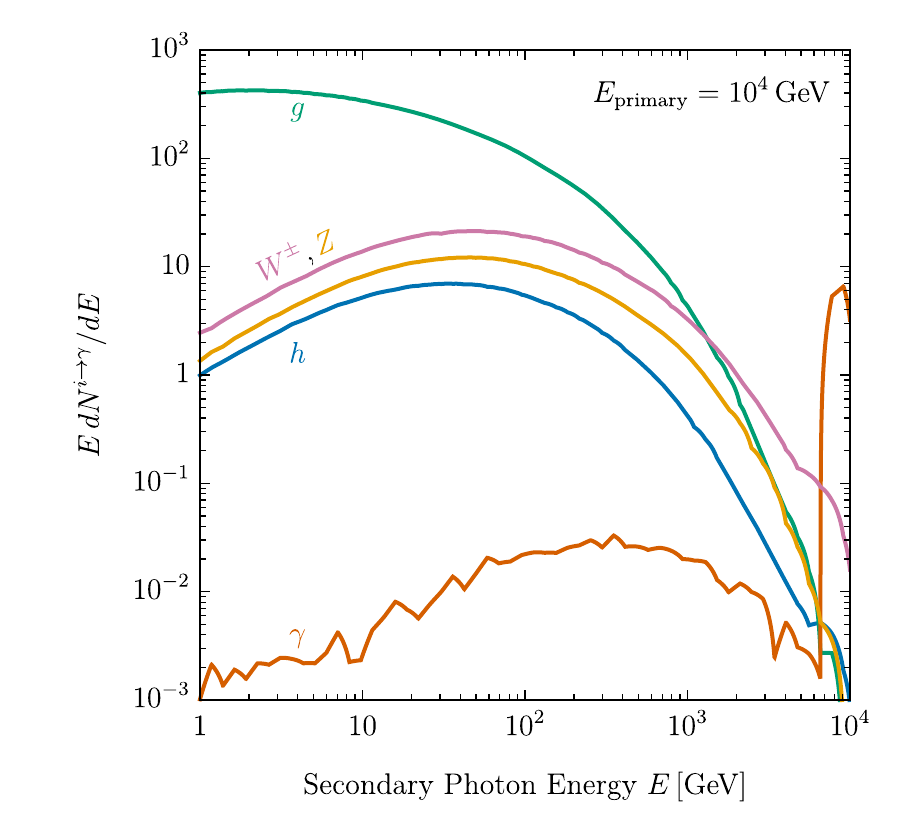}\\
    \includegraphics[width=0.475\textwidth]{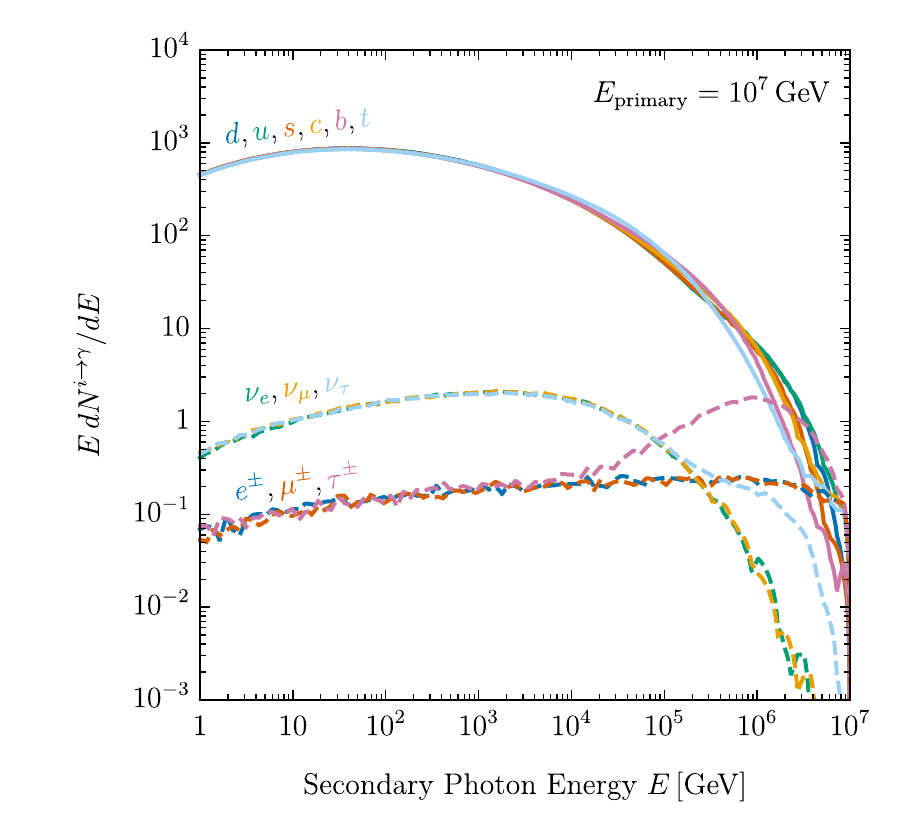}&
    \includegraphics[width=0.475\textwidth]{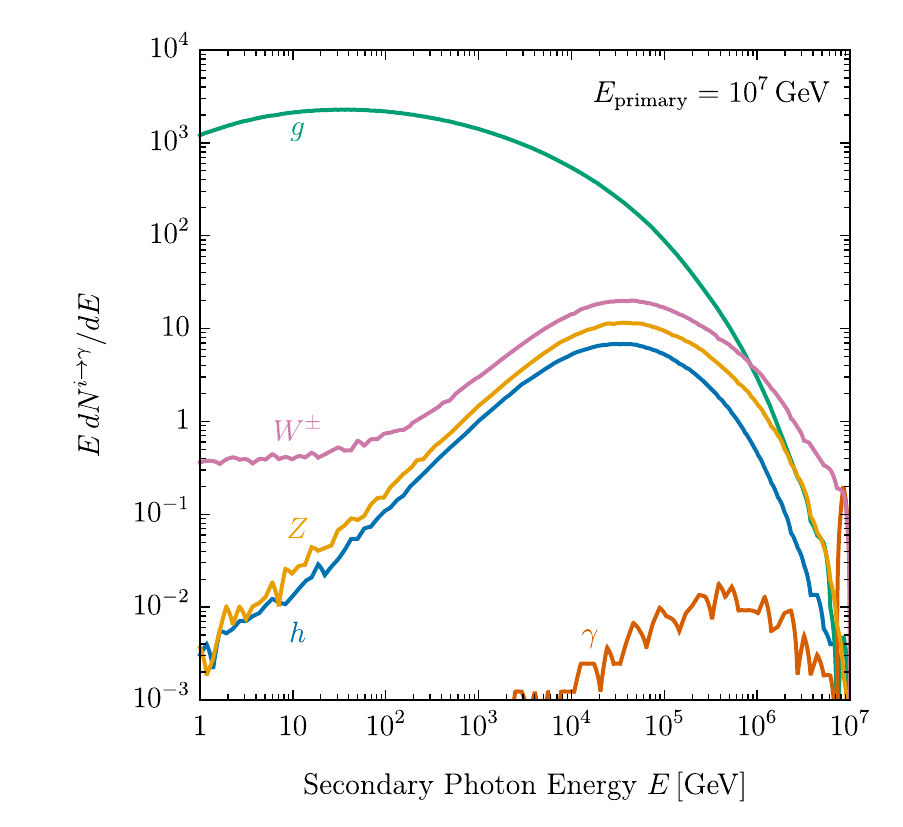}
    \end{tabular}
    \caption{The secondary photon spectra per primary particle for the SM fermions (left) and bosons (right) at a primary energy of $10^4\,\text{GeV}$ (top) and $10^7\,\text{GeV}$ (bottom).  The artefacts are Monte Carlo error for rare processes and do not lead to a significant error in the total secondary spectra.}
    \label{fig:secondaries}
\end{figure}

As well as the primary photons, all SM particles that are directly emitted from the black hole produce secondary photons as they decay, shower and/or hadronise.  This is a complicated process to calculate and we use \texttt{pythia8}~\cite{Sjostrand:2014zea} to compute the secondary photon spectra for each SM particle for a wide range of energies ($1\,\text{GeV}$ to $10^7\,\text{GeV}$).  Since it was developed for a collider environment, \texttt{pythia} loses precision above $10^7\,\text{GeV}$, and it does not compute particle evolution below $1\,\text{GeV}$.   We show the secondary spectra for the SM particles at two primary energies in \cref{fig:secondaries}.  In the top row we show the secondary photon spectra for SM particles of initial energy $10^4\,\text{GeV}$, with fermions on the left and boson on the right.  For fermions, we see that the quarks emit a large number of low energy photons, mostly produced by pion decay in the hadronisation process.  While there is not a great difference between the flavours, the top quark produces more photons at lower energies while the light quarks produce more high energy secondary photons.  The spectra for electrons and muons are approximately flat while taus produce an increased number of mid and high energy photons, due to their hadronic decays.  The neutrinos emit more low and mid energy photons, but significantly fewer high energy photons.  A small enhancement for tau neutrinos can be seen at high energies, as they have a higher chance of producing taus.  The plots are not smooth at low fluxes due to Monte-Carlo error, but since these regions do not significantly contribute to the final flux, the final photon spectra we calculate are smooth.

%-----------------------------------------------------------------
%-----------------------------------------------------------------

For the bosons at $10^4\,\text{GeV}$ (top right panel of \cref{fig:secondaries}), we see that the gluon produces more low energy photons than the quarks, but fewer high energy photons than the lightest quarks.  The $W^\pm$, $Z$ and $h$ produce more low and mid energy photons than the charged leptons, due to their hadronic decays, but fewer high energy photons.  While the photon produces secondary photons via final state radiation, this is subdominant to all the other contributions.  Rather than introduce errors by using the secondary peak $\sim 10^4\,\text{GeV}$, we instead use the primary spectra for the photon and neglect the secondary photon spectrum.  This overestimates the high energy photon flux by at most 5\%.

The lower panels of \cref{fig:secondaries} show the secondary photon spectra for primary particles of energy $10^7\,\text{GeV}$.  Of the fermions, the quarks again give the largest flux, and the differences between the flavours is reduced because the mass effects are less relevant at these high energies.  The neutrinos again produce more low energy photons than the charged leptons, and high energy peaks are seen in the flux from taus and tau neutrinos.  The behaviour for the bosons is similar to that at $10^4\,\text{GeV}$, although we can now see that $W^\pm$ produces more photons at the lowest energies than the neutral $Z$ and $h$.

We similarly use \texttt{pythia8} to generate the secondary spectra for our MSSM benchmark, after creating a spectrum file using \texttt{spheno}~\cite{Porod:2003um,Porod:2011nf} with the $M_h^{125}$ parameter values from ref.~\cite{Bagnaschi:2018ofa} with $M_A = 2\,\text{TeV}$ and $\tan\beta = 20$.  We show the spectra for the individual MSSM particles in \cref{sec:mssm-secondaries}.  Since the decays of the $125\,\text{GeV}$ Higgs boson differ slightly from the SM Higgs boson in this model, we recompute its secondary spectra.  All other SM particles (including the top) are unaffected.  We make the spectra for both the SM and our MSSM benchmark publicly available at \href{https://github.com/PhysicsBaker/SecondarySpectra}{github.com/physicsbaker/secondaryspectra}.

\texttt{Pythia} was designed to model showering and hadronisation at colliders, so focuses on the GeV to the TeV scale.  At higher scales the results become less accurate.  For instance, it does not include triple gauge couplings, which become increasingly important at higher scales.  We compared our \texttt{pythia8} results to \texttt{HDMSpectra}~\cite{Bauer:2020jay}, which is explicitly designed to accurately model showering and hadronisation up to the Planck scale.  We found, as noted in ref.~\cite{Bauer:2020jay}, small corrections to the secondary photon spectra for coloured particles and larger corrections for uncoloured particles, exceeding an order of magnitude at some secondary photon energies.  Nevertheless, we opted to use \texttt{pythia8} for two reasons.  Firstly, at energies greater than a GeV, secondary photon emission is dominated by the contribution from coloured particles, which is well modelled by \texttt{pythia8}.  Secondly, \texttt{pythia8} can be used to compute the secondary spectra for BSM particles, such as those in the MSSM, while \texttt{HDMSpectra} cannot.  We thus chose to compute both secondary spectra on the same footing. In models where \texttt{pythia} is not expected to give sufficient accuracy at high scales, the \texttt{pythia} plug-in \texttt{vincia}~\cite{Fischer:2016vfv} could be used.

Now we have calculated the secondary spectra for primary particles of a given energy, $d N^{i\to\gamma}/dE_\text{s}(E_\text{p},E_\text{s})$, we can integrate this against the primary spectra, $d^2 N^i_\text{p}/dt dE_\text{p}(M,E_\text{p})$, to find the secondary spectra for a Schwarzschild black hole of a given mass.  Summing over all emitted particle species we find the total secondary photon flux:
\begin{align}
    \label{eq:d2NdEdt-secondaries}
    \frac{d^2 N^{\gamma}_\text{s}}{dt dE_\text{s}} =&\, 
    \sum_{i} \int_0^\infty
    \frac{d^2 N^i_\text{p}}{dt dE_\text{p}}(M,E_\text{p}) 
    \frac{d N^{i\to\gamma}}{ dE_\text{s}}(E_\text{p},E_\text{s})
    dE_\text{p} \,.
\end{align}
As discussed above we use a delta function for $d N^{\gamma\to\gamma}/dE_\text{s}$. 

%-----------------------------------------------------------------
\begin{figure}
    \centering
    \begin{tabular}{cc}
    \includegraphics[width=0.475\textwidth]{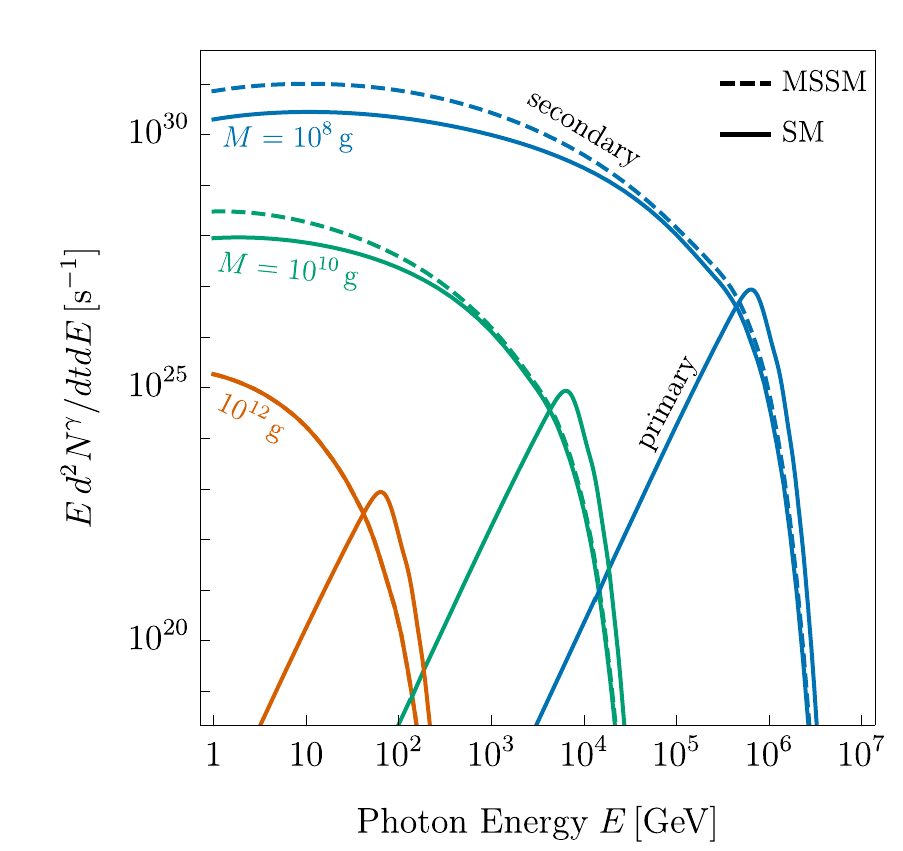}&
    \includegraphics[width=0.475\textwidth]{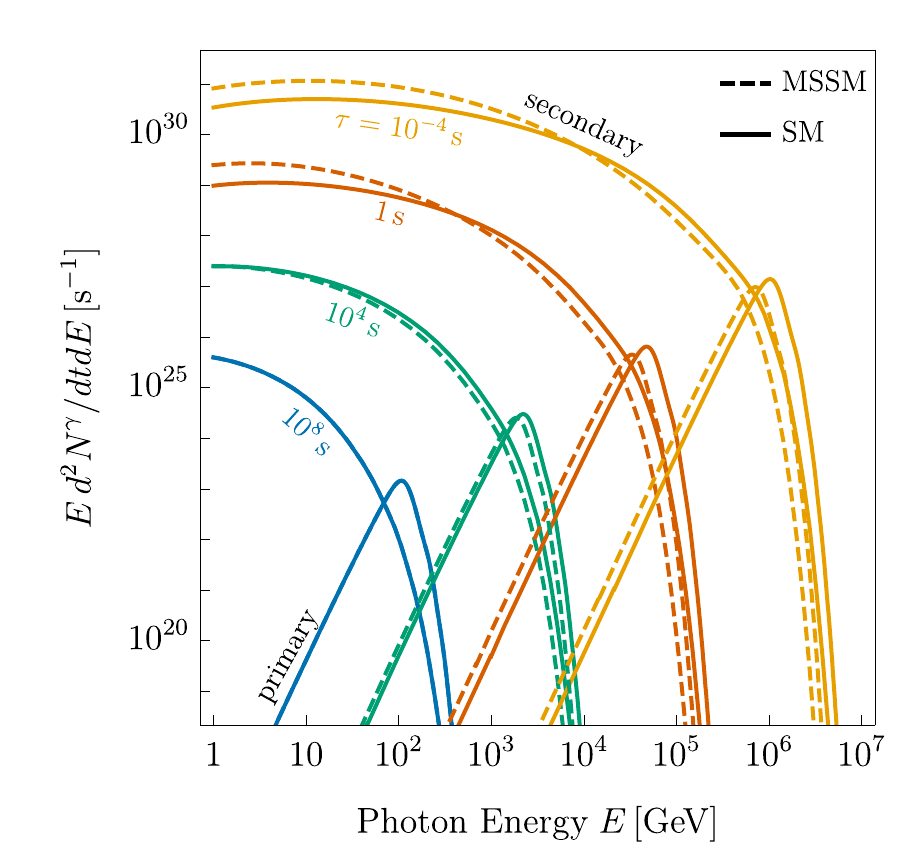}\\
    \includegraphics[width=0.475\textwidth]{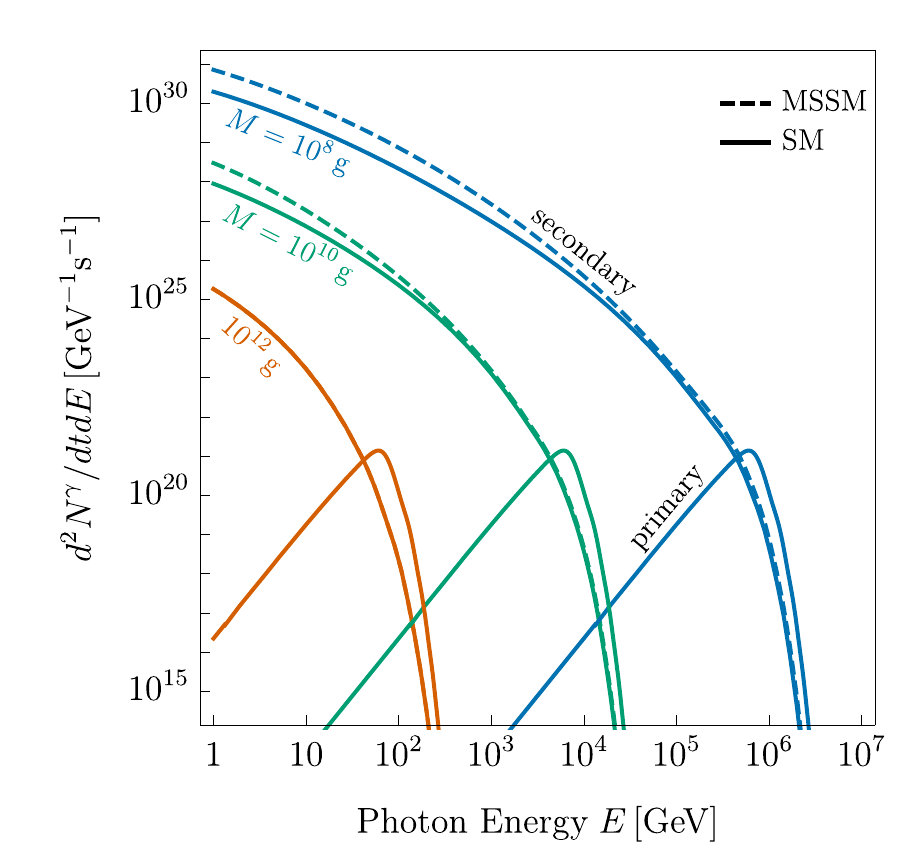}&
    \includegraphics[width=0.475\textwidth]{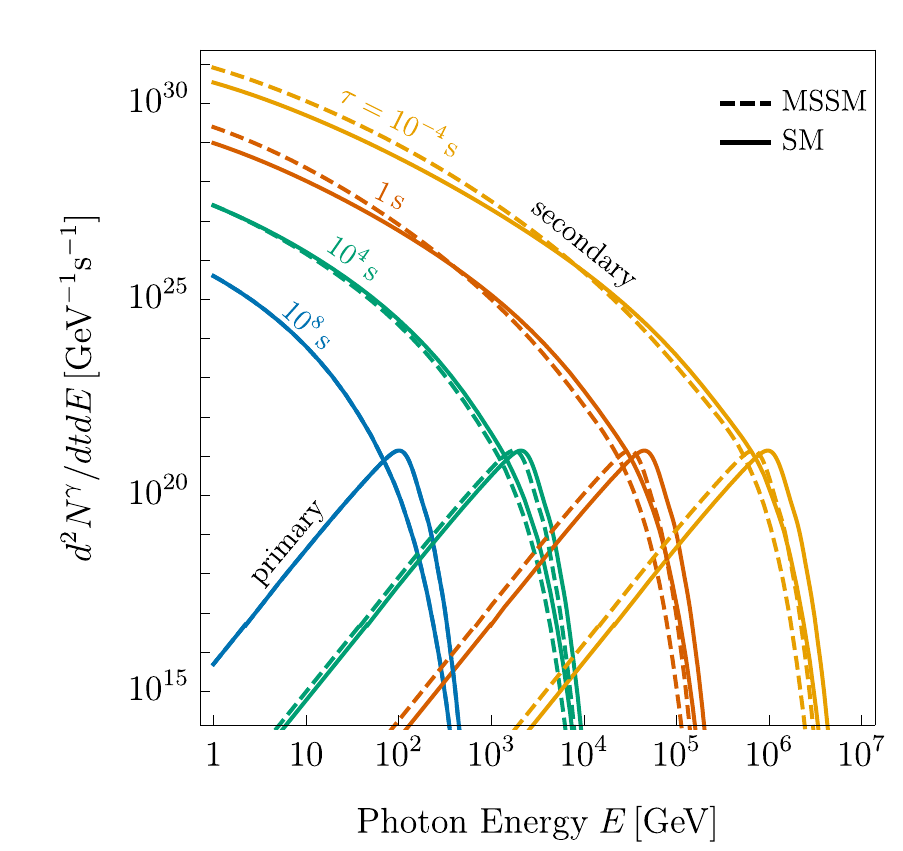}
    \end{tabular}
    \caption{Primary and secondary photon spectra for the SM and MSSM for $M = 10^{12}, 10^{10}, 10^8\,\text{g}$ (top left) and $\tau = 10^8, 10^4, 10^0, 10^{-4}\,\text{s}$ (top right).  In the lower panels we show the same but without multiplying the flux, $d^2N^\gamma/dtdE$, by the energy, $E$.
    }
    \label{fig:primary-secondary}
\end{figure}
%-----------------------------------------------------------------

The secondary spectra for all particles excluding the photon (which from now on we will simply call the secondary spectra), and the primary photon spectra, are shown in \cref{fig:primary-secondary}.  In the top left panel we show $E\, d^2 N^\gamma / dtdE$ for black holes of a given mass.  For most of the BSM models we consider, there are no new particles which produce a significant number of secondary photons.  As such, these models will have the same spectra as in the SM scenario.  However, in the MSSM benchmark there are many new particles which produce secondary photons, as discussed in \cref{sec:mssm-secondaries}.  We see from the dashed lines in \cref{fig:primary-secondary} (top left) that at black holes masses of $10^8\,\text{g}$ and $10^{10}\,\text{g}$ there are significantly more secondary photons produced, particularly at low energies.  However, at $10^{12}\,\text{g}$ the secondary spectrum is identical to the SM case.  This is because a $10^{12}\,\text{g}$ black hole has a temperature around $10\,\text{GeV}$, so it is not yet hot enough to emit a significant number of MSSM particles, which all have masses above $\sim 1\,\text{TeV}$.  Note that the secondary spectra are now smooth and do not suffer from the Monte-Carlo errors seen at low fluxes in the individual channels.  We also make these spectra publicly available at \href{https://github.com/PhysicsBaker/SecondarySpectra}{github.com/physicsbaker/secondaryspectra}.

In \cref{fig:primary-secondary} (top right) we show the primary and secondary photon spectra at fixed times, given by $\tau$, for the SM and MSSM scenarios.  We see that at $\tau = 10^8\,\text{s}$ the spectra are identical.  This is because at this $\tau$ the black hole is not yet hot enough to emit the heavy MSSM particles.  At $\tau \sim 10^4\,\text{s}$ the black hole becomes hot enough to emit a significant flux of heavy MSSM particles and, at the same $\tau$, the black hole in the MSSM scenario is cooler than the SM case (since the extra degrees of freedom mean that the subsequent mass loss of the black hole will be faster).  We see that from this time on the peak of the MSSM primary photon spectrum occurs at lower energies than in the SM case.  On the other hand, the MSSM scenario still results in a greater number of low energy secondary photons.  These two effects pull in opposite directions and it is not \textit{a priori} clear whether the MSSM scenario results in more or less photons than the SM scenario overall.  Integrating these spectra, we find that the SM emits more total photons above $10^4\,\text{GeV}$ than our benchmark MSSM scenario ($3.6\times10^{27}$ to $1.5\times10^{27}$) in the time window $10^{-6}\,\text{s} < \tau < 10^4\,\text{s}$, and more total photons above $10^2\,\text{GeV}$ in the same time window ($7.0\times10^{30}$ to $5.0\times10^{30}$).  This is in contrast to the conclusion reached in ref.~\cite{Ukwatta:2015iba}, which added a single squark to the SM and neglected the change in the mass loss rate.  In that case, more photons were expected to be released in the final burst than in the SM scenario.

In the lower panels of \cref{fig:primary-secondary} we show the same information, without multiplying the flux, $d^2N^\gamma/dtdE$, by the energy, $E$.  We see that the peak of the primary spectra no longer rises as the black hole heats up.  Even though more photons are emitted per second, this is accounted for by the changing scale of the $x$-axis.  This presentation gives a clearer connection to the features seen in \cref{fig:contour-1,fig:contour-2}, discussed below.

%=============================================================================
\section{Distinguishing the SM From BSM Scenarios}
\label{sec:sm-bsm}
%=============================================================================

If an evaporating black hole is detected, some of the key questions to answer are (i) is there any evidence of new degrees of freedom in nature and (ii) if so, do those new degrees of freedom produce secondary photons.  These two questions would give us very important information on (i) the fundamental particles present in nature and (ii) how best to look for them in other experiments.  To best answer these questions an alternative presentation to \cref{fig:primary-secondary} is perhaps preferable.

%-----------------------------------------------------------------------------
\subsection{Models without Extra Secondary Photons}
%-----------------------------------------------------------------------------

%-----------------------------------------------------------------
\begin{figure}
    \centering
    \begin{tabular}{cc}
        \includegraphics[width=0.475\textwidth]{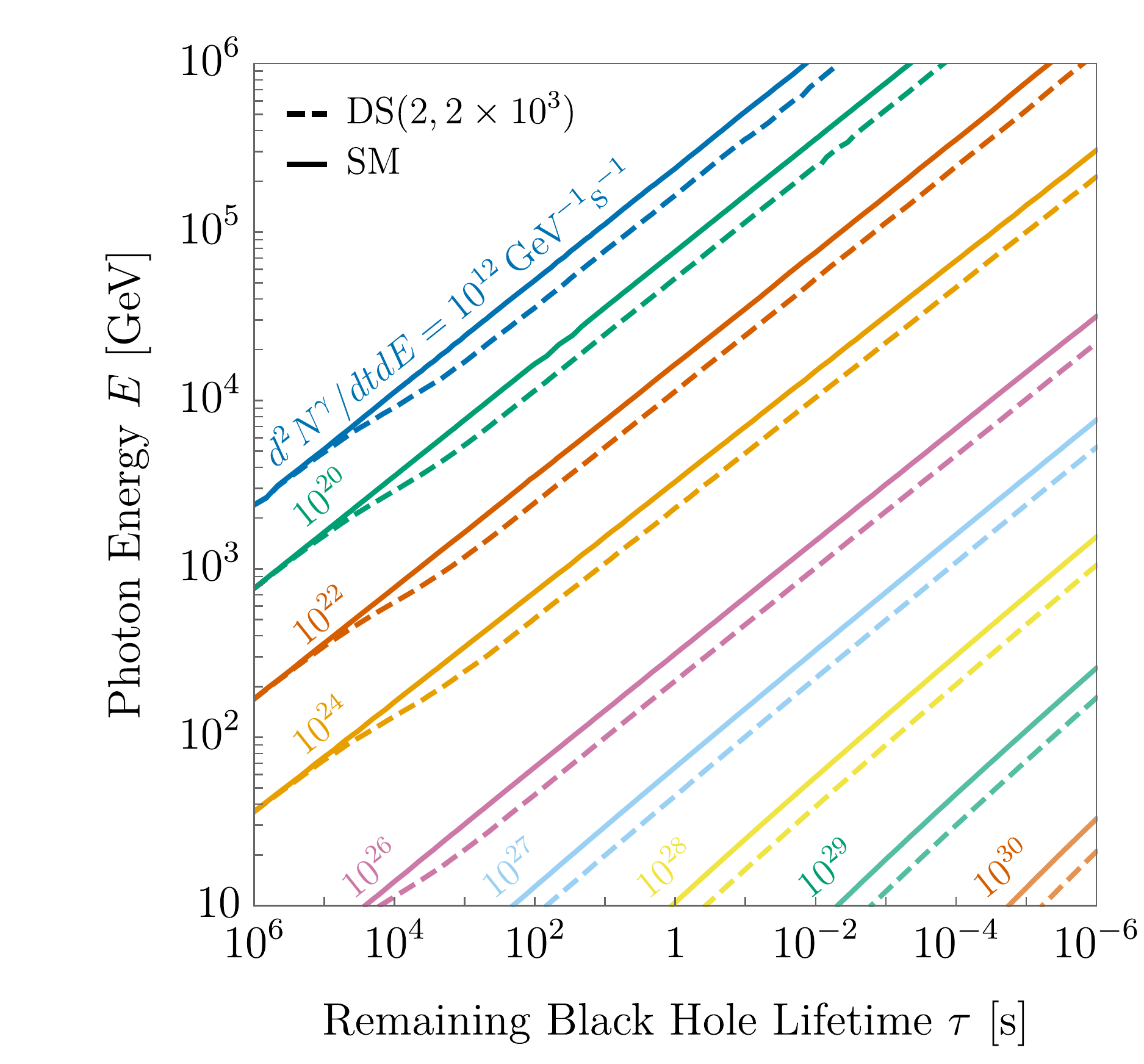}&
    \includegraphics[width=0.475\textwidth]{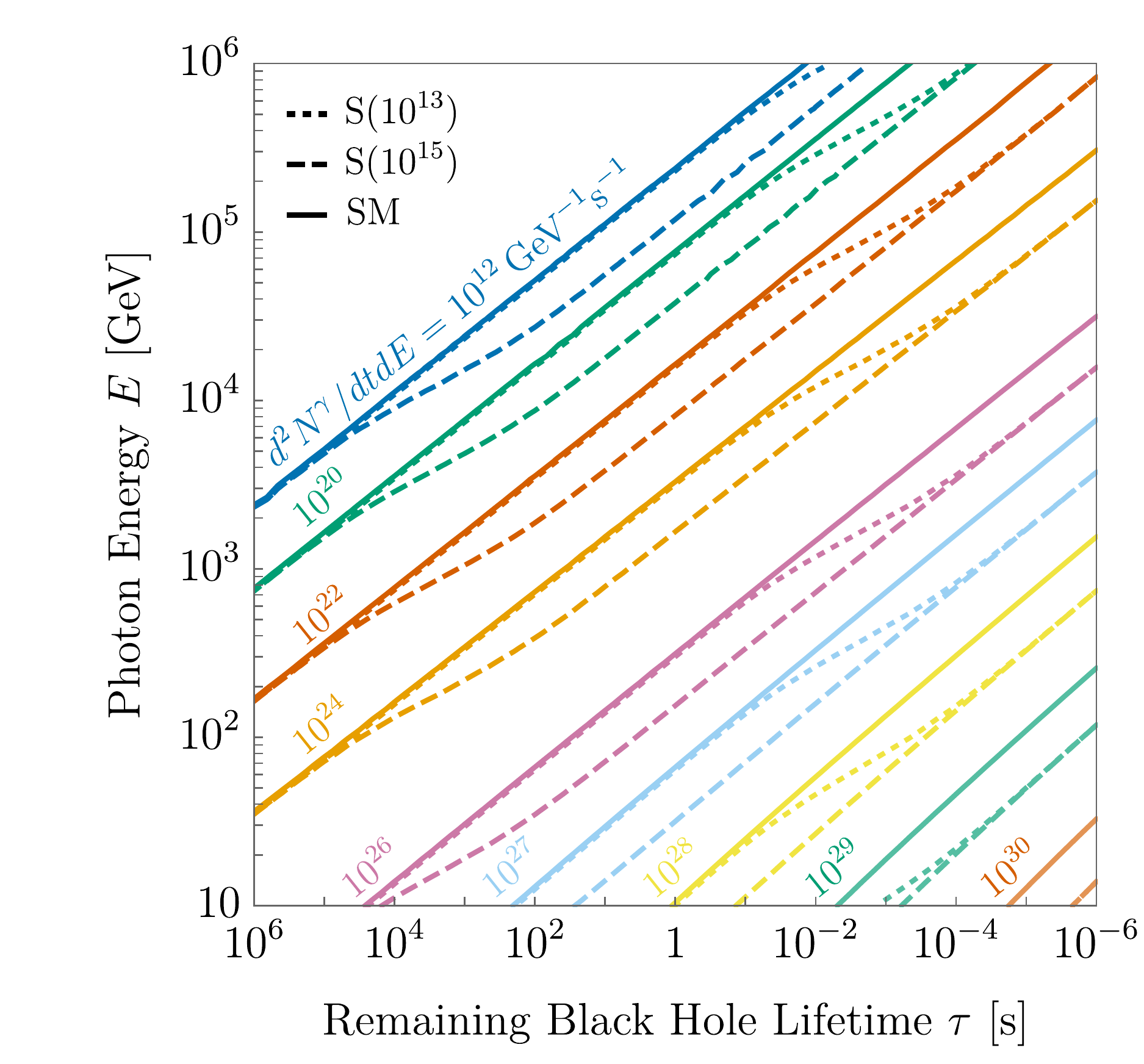}
    \end{tabular}
    \caption{Contours of the total photon flux as a function of remaining time, $\tau$, and photon energy, $E$.  We show the SM expectation with solid contours and in dashed contours we show the DS$(2,2\times10^3\,\text{GeV})$ benchmark (left), the string inspired benchmarks (right).
}
    \label{fig:contour-1}
\end{figure}
%-----------------------------------------------------------------

As noted in \cref{sec:particle-emission}, models which do not produce significant numbers of secondary photons can still alter the signature of an evaporating black hole, by changing its evaporation rate.  In the left panel of \cref{fig:contour-1} we show the total photon spectra for the SM and the DS$(2,2\times10^3\,\text{GeV})$ benchmark  as a contour plot in time and photon energy.  We see that before $\tau \sim 10^5\,\text{s}$ the two spectra are identical.  This is the period when the black hole is not yet hot enough to emit the dark sector particles, which have a mass scale of $2\times10^3\,\text{GeV}$.  At smaller $\tau$, these new degrees of freedom begin to be emitted, leading to the black hole to lose mass at a higher rate.  This leads to it being substantially cooler than the SM scenario for $\tau \lesssim 10^5\,\text{s}$, and a reduction in the photon flux across the entire range of energies. The DS$(1,10^4\,\text{GeV})$ benchmark similarly deviates from the SM at $\tau \sim 10^3\,\text{s}$, when the temperature increases above the mass threshold,  and reduces the photon flux less than DS$(2,2\times 10^3\,\text{GeV})$, as there is now only one copy of the SM dof.  DS$(10,10^2\,\text{GeV})$ decreases the photon flux even more significantly from $\tau \sim 10^9\,\text{s}$ up to the end of the evaporation as it introduces many more new dof.

\Cref{fig:contour-1} (right) shows a similar plot for the SM and the string inspired benchmarks.  We see that the deviations now start at $\tau \sim 10^5\, (10^{-1})\,\text{s}$ for $\mathcal{V} = 10^{15}\, (10^{13})$ and that the reduction in the photon flux is larger than the DS$(2,2\times10^3\,\text{GeV})$ benchmark.  This is because the string inspired benchmarks introduce more new degrees of freedom than the DS$(2,2\times10^3\,\text{GeV})$ benchmark, and the $\mathcal{V} = 10^{15}\,(10^{13})$ scenario introduces them at a scale of $2\times10^3\, (2\times10^5)\,\text{GeV}$.  We can imagine that the results are analogous for other models where no extra photons are produced: there will be a reduction in the flux at energies where new fundamental particles begin to be emitted. In the $N$naturalness benchmark the photon flux is reduced by many orders of magnitude compared to the SM.

%-----------------------------------------------------------------------------
\subsection{Models with Extra Secondary Photons}
%-----------------------------------------------------------------------------

%-----------------------------------------------------------------
\begin{figure}
    \centering
    \begin{tabular}{cc}
    \includegraphics[width=0.475\textwidth]{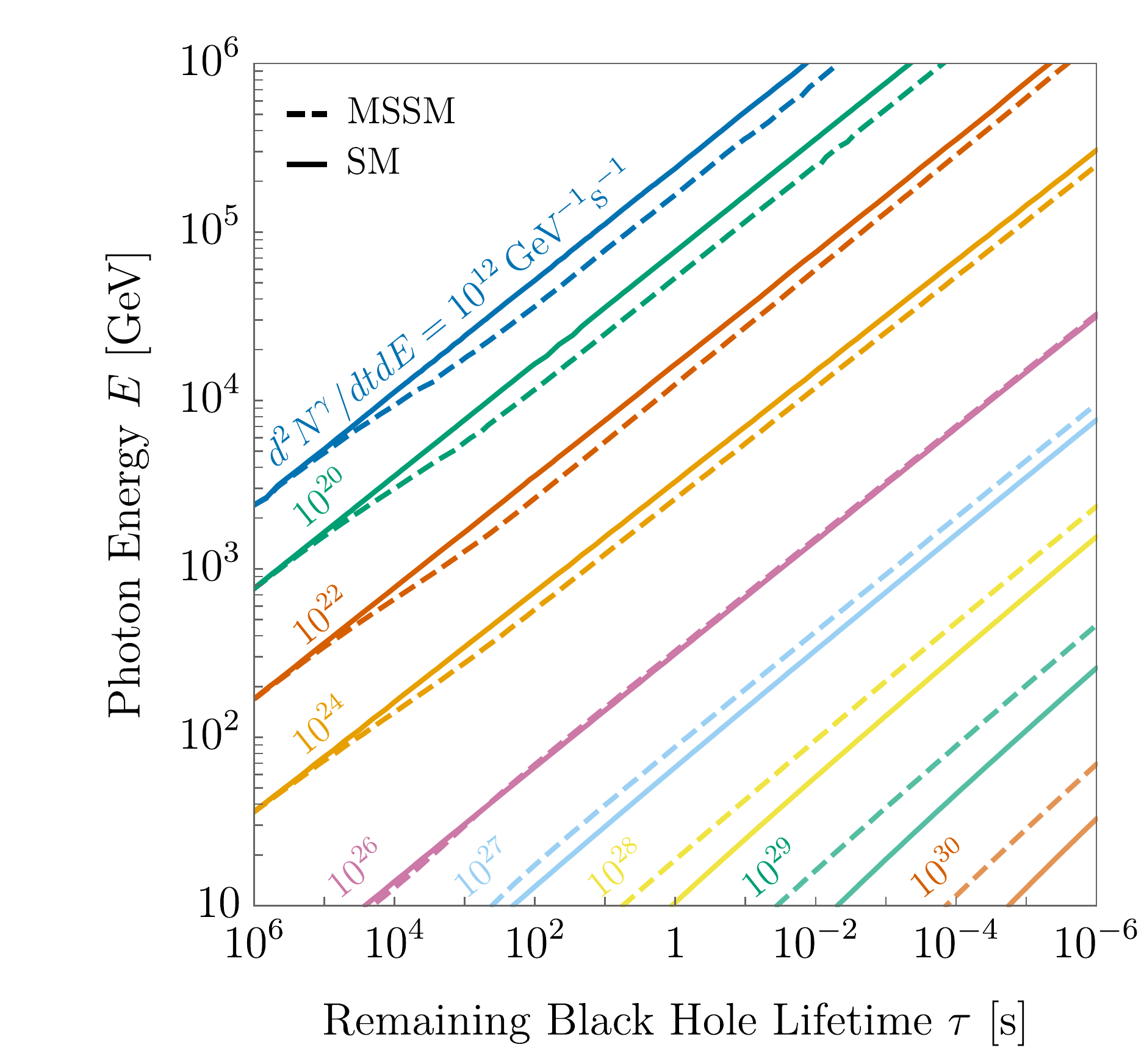}&
    \includegraphics[width=0.475\textwidth]{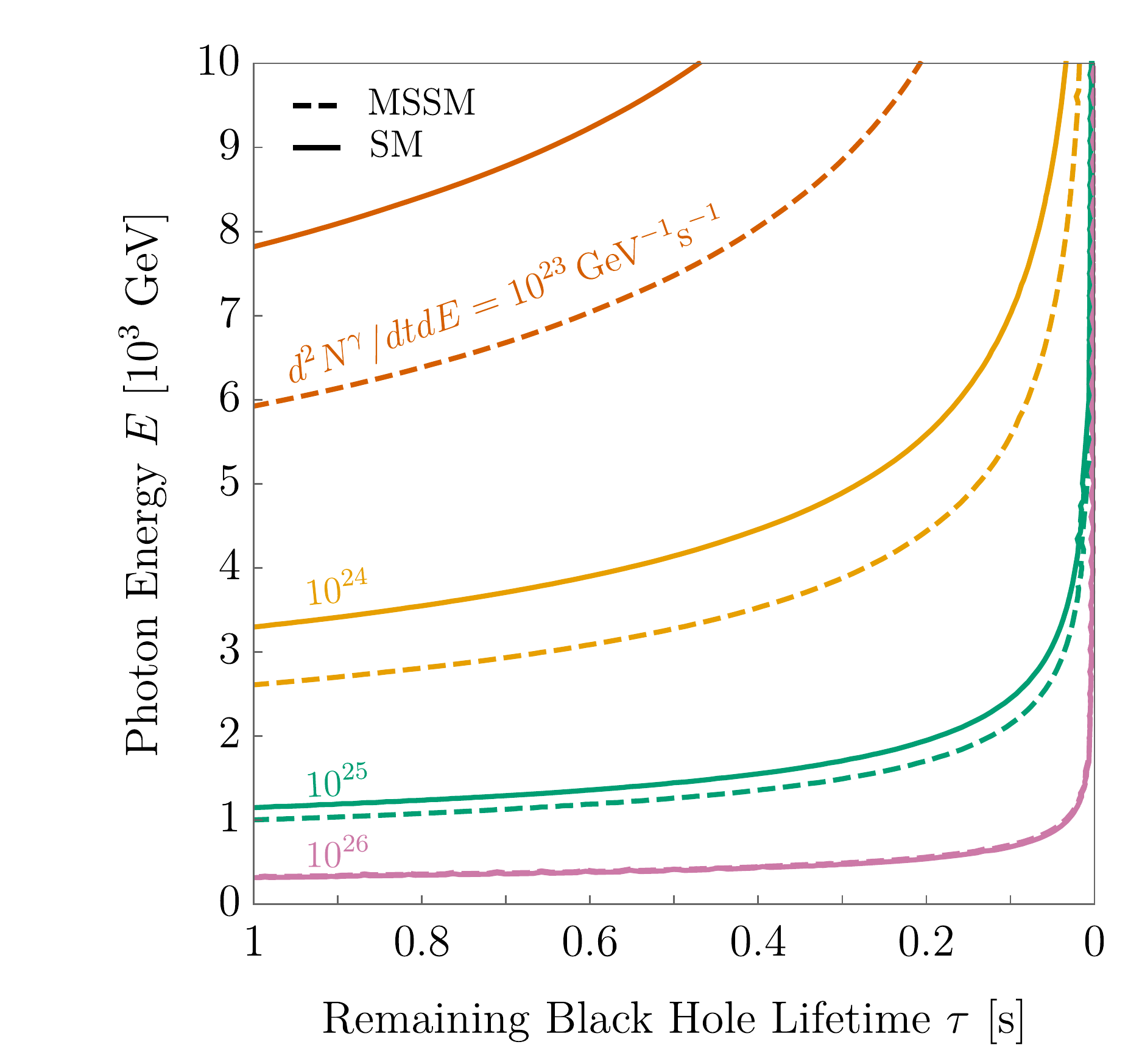}
    \end{tabular}
    \caption{Contours of the total photon flux as a function of remaining time, $\tau$, and photon energy, $E$.  We show the SM expectation with solid contours and in dashed contours we show the MSSM benchmark.  We show the flux on a logarithmic scale (left) and a linear scale (right).
}
    \label{fig:contour-2}
\end{figure}
%-----------------------------------------------------------------

Other models, such as our MSSM benchmark, contain new particles which produce a significant secondary photon spectrum.  In \cref{fig:contour-2} we show contour plots in time and photon energy for the MSSM benchmark.  In the left panel we see that there is no deviation at $\tau \gtrsim 10^5\,\text{s}$, again because for these times the black hole is too cool to produce the MSSM particles, which have a mass greater than $\sim 10^3\,\text{GeV}$.  For $\tau \lesssim 10^5\,\text{s}$ we see that at the highest energies there is a reduction in the photon flux, but that at the lowest energies there is an increase.  This is the result of two competing effects: the new degrees of freedom produce many low energy photons, but they also speed up the mass loss of the black hole, leading to lower temperatures before the end of explosion.  We see that the two effects approximately cancel out along the $d^2N^\gamma /dtdE = 10^{26}\,\text{GeV}^{-1}\text{s}^{-1}$ contour.  As noted at the end of \cref{sec:photon-spectra}, the SM produces more total photons above $100\,\text{GeV}$.  As a comparison, we see from \cref{fig:alpha} that the MSSM benchmark and the DS$(2, \times 10^3\,\text{GeV})$ benchmark have very similar $\alpha(M)$'s, so will have very similar time evolution of the black hole mass.  Comparing the left panel of \cref{fig:contour-1}and the left panel of \cref{fig:contour-2} we see that the flux of the highest energy photons is very similar between these models, but that the MSSM has a greater flux for energies lower than $d^2N^\gamma/dtdE = 10^{22}\,\text{GeV}^{-1}\text{s}^{-1}$, and that this behaviour is independent of time for $\tau \gtrsim 10^5\,\text{s}$.

It should also be noted that the overall effect, in all models, is not as modest as it appears on a log plot.  In the right panel of \cref{fig:contour-2} we show a portion of the spectrum for the MSSM benchmark on a linear plot, where we see that the effect is significant.

%-----------------------------------------------------------------------------
\subsection{Analysis Strategies}
%-----------------------------------------------------------------------------

We have seen in the previous sections that the photon spectra, as a function of time, encodes crucial information about the fundamental particles present in nature.  New particles beyond the SM change both the mass evolution with time, and the secondary photon spectra at a given mass.  Huge numbers of new particles or changes to the nature of space-time could also mean that the fundamental Planck scale is much lower than currently thought, leading to a much earlier evaporation end-point.

However, whether we can probe nature via an evaporating black hole depends on whether one will be observed.  This depends both on the experiments that are built and run, and on the population of exploding black holes within $\sim 1\,\text{pc}$.  While the direct limits on the population are not discouraging (the strongest direct limit is $3400 \,\text{pc}^{-3}\,\text{yr}^{-1}$~\cite{Albert:2019qxd}, which corresponds to a probability of the HAWC experiment observing at least one event in the next five years at a distance less than $0.05$ $(0.01)\,\text{pc}$ of $\sim 83\%$ $(1.4\%)$), indirect limits due to galactic and extra-galactic $100\,\text{MeV}$ photons and anti-protons seem significantly stronger~\cite{1976ApJ...206....1P,1977ApJ...212..224P,Lehoucq:2009ge,Abe:2011nx,Carr:2020gox,Auffinger:2022khh}.  These limits, however, rely on a variety of assumptions, such as degree of clustering~\cite{1976ApJ...206....1P,1977ApJ...212..224P,Carr:2020gox}, anti-proton production and propagation models~\cite{Abeysekara:2013ska}, details of emission near the QCD scale ($\Lambda_\text{QCD} \sim 200\,\text{MeV}$)~\cite{MacGibbon:1990zk,MacGibbon:1991tj,Coogan:2020tuf,Carr:2020gox,Arbey:2021mbl}, and the presence of new degrees of freedom.  Indeed, we see from \cref{fig:Ep-tau} that in an $N$naturalness framework, an evaporating black hole would only have a temperature around $100\,\text{MeV}$ in the last $100\,\text{s}$ of its life.  This demonstrates that, in principle, new fundamental particles beyond the SM can significantly alter the signature from an exploding black hole, and it is clear that this would directly impact the current direct and indirect limits.  As such, we take the view expressed in ref.~\cite{Carr:2020gox} and, accepting that our understanding is currently incomplete, focus on the empirical aspects of an observation.  We therefore briefly turn to possible analysis strategies for BSM physics following an observation.

\begin{figure}
    \centering
    \begin{tabular}{cc}
    \includegraphics[width=0.475\textwidth]{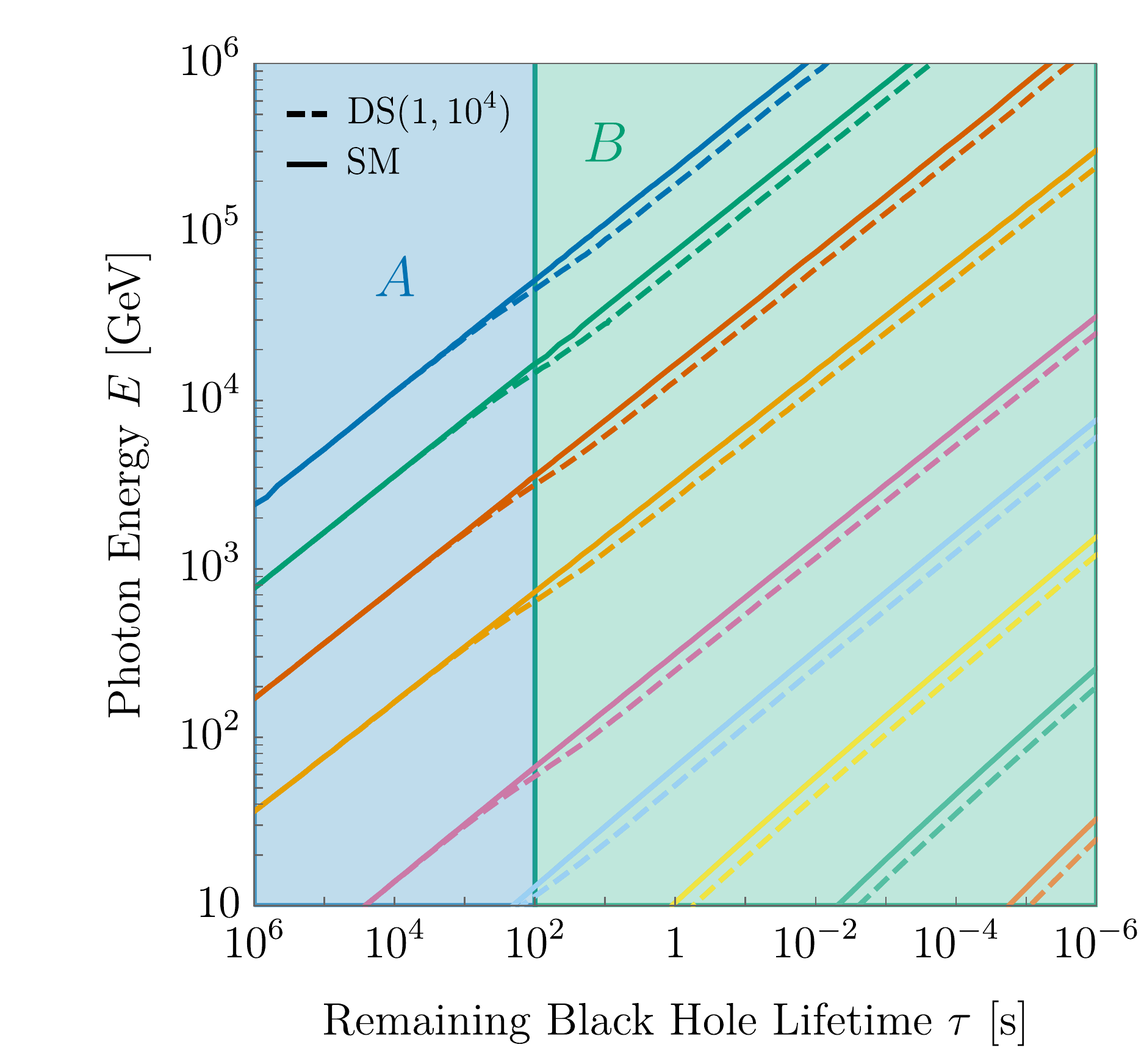}&
    \includegraphics[width=0.475\textwidth]{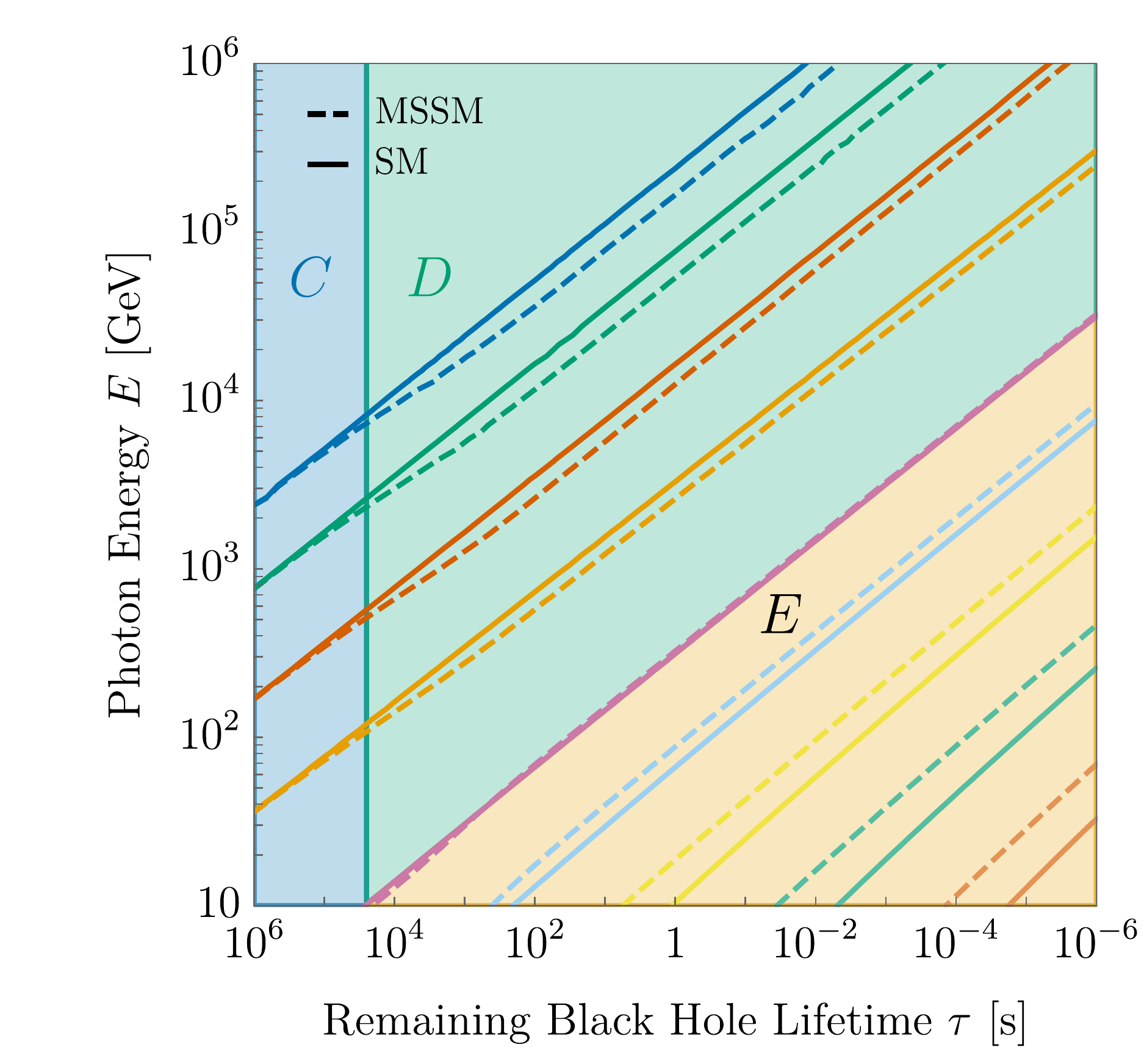}
    \end{tabular}
    \caption{Possible binning regions for models with new particles which only interact weakly with photons (left), in this case the DS$(1,10^4)$ benchmark with the same contours as \cref{fig:contour-1} (left), and models which produce a significant number of secondary photons (right), in this case our MSSM benchmark.}
    \label{fig:analyses}
\end{figure}

The optimal experimental analysis following an observation would depend on the total number of photons observed, as well as the energy and time resolutions and thresholds, and the backgrounds of the particular experiment.  However, prior to an observation we can make some general suggestions, assuming that the signal is broadly SM-like.  We can expect that relatively few photons would be received in a first observation, perhaps $\mathcal{O}(10-1000)$, since there is a greater probability of observing an exploding black hole from further away if they are uniformly distributed in the region $\sim 1 \,\text{pc}$ away from the Earth.  As such, it may be preferable to bin the events in just a few bins, to reduce statistical errors.  An analysis strategy for models which do not produce significant numbers of secondary photons, such as those shown in \cref{fig:contour-1}, could be to fix a mass scale of interest, $\widetilde\Lambda$, find the  $\tau$ in the SM scenario where $T(\tau)\sim\widetilde\Lambda$, and then calculate the ratio of photons observed before and after that $\tau$ value.  That is, to use the ratio of photons observed in region $A$ to those observed in region $B$ of \cref{fig:analyses} (left) as the key discriminating observable.

%-----------------------------------------------------------------
%-----------------------------------------------------------------

For analyses probing BSM models which produce a significant number of secondary photons, it could be beneficial to separate the region of low energy photons along a contour of constant photon flux in the SM, as shown in \cref{fig:analyses} (right).  That is, to use the ratio of the number of photons observed in regions $C$ to $D$, the ratio of $C$ to $E$ and the ratio of $D$ to $E$.  In this way the region where the BSM model predicts fewer photons is separated from the region where it predicts more photons.  Since the flux where the SM contour matches the BSM contour will depend on the details of the BSM model, separate analyses could target different assumptions.

%-----------------------------------------------------------------------------
\subsection{Impact of an Observation}
%-----------------------------------------------------------------------------

An observation of an evaporating black hole would have a profound impact on our understanding of a wide range of BSM theories, and of the fundamental nature of the universe.  Firstly, we see that models which introduce a huge number of light particles would radically alter the signature from an evaporating black hole.  The $N$naturalness benchmark would lose mass at such a huge rate that it would not produce $100\,\text{MeV}$ photons until its last $100\,\text{s}$, and would only produce photons above $10\,\text{GeV}$ in the last $10^{-5}\,\text{s}$.  This scenario would call for a radically different analysis strategy than the ones currently employed and would require a complete re-evaluation of the existing limits.  Conversely, observation of a reasonably SM-like evaporation process would immediately and conclusively rule out the whole class of models with a large number of light or massless particles.

Secondly, as noted above and in \cref{sec:models}, models which change the nature of space-time, such as extra-dimensional models, or which drastically reduce the fundamental Planck scale, as would be the case if there were $10^{32}$ copies of the SM particles~\cite{Dvali:2007hz,Dvali:2007wp,Calmet:2008tn,Dvali:2009ne}, would lead to an endpoint of the explosion at much lower energies than usually envisioned.  These scenarios would similarly drastically alter the photon signature and would require new analysis strategies.  In this case, an observation of a reasonably SM-like evaporation process would imply that the fundamental Planck scale is at least as high as the highest energy photon observed, potentially above $10^7\,\text{GeV}$.  This is orders of magnitude larger than the current limits and would be very general, applying to a very broad range of models (e.g., both extra-dimensional models and those that lower the Planck scale with many new heavy degrees of freedom).

Finally, for more commonly studied extensions of the SM, such as dark sectors and the MSSM, we have seen how both the evaporation rate and the secondary photon spectra can be affected.  Models containing new particles which only couple weakly to the photon all have a similar impact on the observational signature: the black hole evaporation is sped up once its temperature reaches the particle mass, leading the black hole remaining cooler for longer before the end of its explosion.  The increase in evaporation rate is greater for particles of lower spin and more internal degrees of freedom.  While we have focused on benchmarks with a common mass scale, more complicated behaviour is expected when there are several new particles at different mass scales.  Nonetheless, the behaviour will be similar: the evaporation rate will speed up every time a new particle threshold is crossed.  If an evaporating black hole signature which differs from the SM expectation is seen, then it would tell us where these thresholds appear, providing information on the mass and spin of new fundamental particles which must be present in nature.  Conversely, observation of a SM-like evaporation would put stringent limits on the possible number of new degrees of freedom present in nature.  Limits of this kind can not be placed in any other way.

While models with new particles which couple strongly to the photon, like the MSSM, produce more photons at a given temperature, we have seen that in the case of our MSSM benchmark fewer photons are produced overall.  This is because the black hole stays cooler for longer, leading to a lower total flux.  We may expect this behaviour to hold in many other models, since the MSSM has a large number of coloured particles which copiously produce secondary photons.  However, we have not yet determined that this is the case for all MSSM parameter points, or for all models in general.  Nonetheless, we have identified possible strategies for extracting information about the new particles masses, using the highest energy photons, and about whether new fundamental particles interact significantly with photons, using the lower energy photons.  Observation of a non-SM-like signal would provide crucial information on the scale and nature of, e.g., new supersymmetric particles.

Conversely, we see that BSM physics can significantly alter the signature of an evaporating black hole.  There is therefore the possibility that existing gamma-ray bursts of unknown origin could actually be the signatures of evaporating black holes, but that the correct particle models are not used to interpret them.  This is a very interesting future direction that may be fruitful to investigate.

%=============================================================================
\section{Conclusions}
\label{sec:conclusions}
%=============================================================================

The observation of an evaporating black hole would provide definitive information on the elementary particles present in nature. It could discover or exclude widely studied models of BSM physics, particularly those with a large number of new degrees of freedom.  In many cases, the models cannot, or are very difficult, to probe in any other way.  In this work we have surveyed a broad range of motivated and widely studied scenarios beyond the standard model of particle physics, discussing in broad terms how these models would affect black hole evaporation.  The models that would have the largest impact are those with a large number of new degrees of freedom, such as supersymmetry, large $N$ solutions to the hierarchy problem, some string inspired models and some dark sector models, as well as those that drastically reduce the fundamental Planck scale, such as large extra-dimensional models and some large $N$ models.  To study the first class of models in greater detail, we defined representative benchmark scenarios.

Since photons provide the most promising experimental avenue for evaporating black hole searches, we calculated the primary and secondary photon spectra as a function of black hole mass and as a function of time for the benchmark scenarios.  Models with many new degrees of freedom below 100 MeV significantly alter the evolution of a primordial black hole from production to today, while models which only introduce new heavy degrees of freedom (above the TeV scale) only impact the final $10^5\,\text{s}$.  Models with new degrees of freedom that produce secondary photons, as they decay, shower and/or hadronise, produce a larger flux of low energy photons, but we found that for the MSSM benchmark scenario the black hole still emits less photons above $100\,\text{GeV}$ overall, since the new degrees of freedom lead to a faster evaporation rate so the black hole stays cooler for longer.

Finally, we presented the photon spectra in the time-energy plane.  After discussing some general aspects relating to experimental observation, we suggest some methods which use this presentation to identify discriminating observables and discuss the profound impact of an potential observation.

%=============================================================================
\section*{Acknowledgements}
%=============================================================================

It is a pleasure to thank Martin Bauer for discussions on extra-dimensional models, Veronica Guidetti for useful discussions on string phenomenology, Joachim Kopp and Raymond Volkas for comments on the manuscript, Werner Porod for advice on using \texttt{SPheno}, and Peter Skands for advice on using \texttt{Pythia}.
We would also like to thank the IPPP at Durham University and the Mainz Institute for Theoretical Physics (MITP) of the Cluster of Excellence PRISMA+ (Project ID 39083149) for their hospitality and partial support during this work.
This work was supported by the Australian Government through the Australian Research Council Centre of Excellence for Dark Matter Particle Physics (CDM, CE200100008).

%-----------------------------------------------------------------------------
\appendix
\section{Secondary Photon Spectra for the MSSM Benchmark Model}
\label{sec:mssm-secondaries}
%-----------------------------------------------------------------------------

%-----------------------------------------------------------------
\begin{figure}
    \centering
    \begin{tabular}{cc}
    \includegraphics[width=0.475\textwidth]{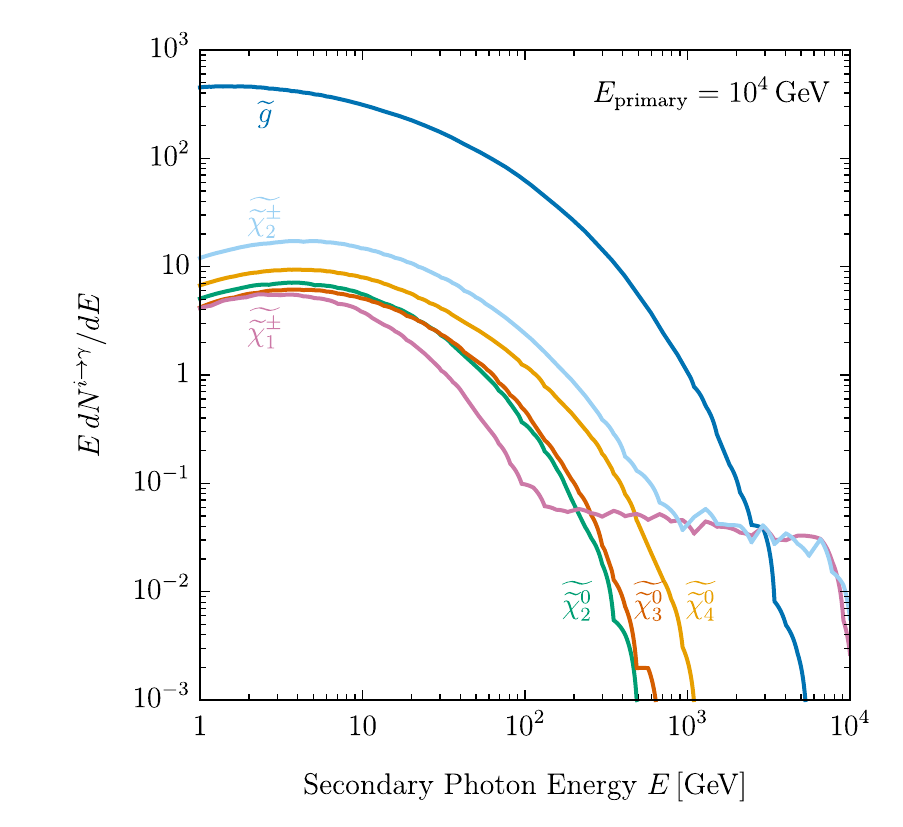}&
    \includegraphics[width=0.475\textwidth]{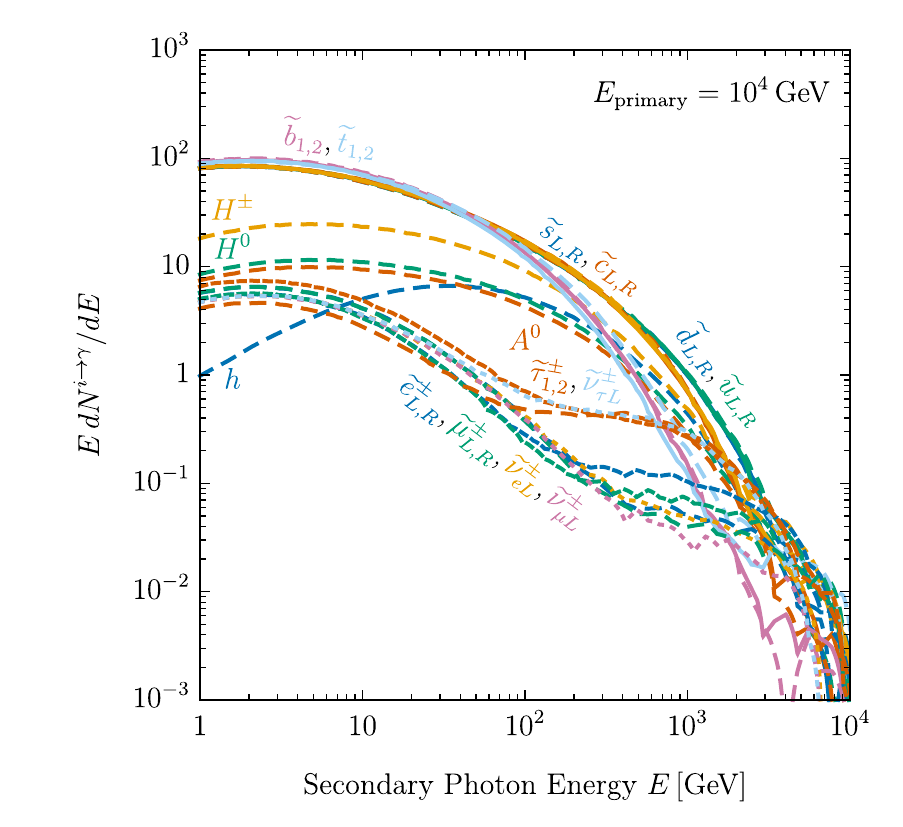}\\
    \includegraphics[width=0.475\textwidth]{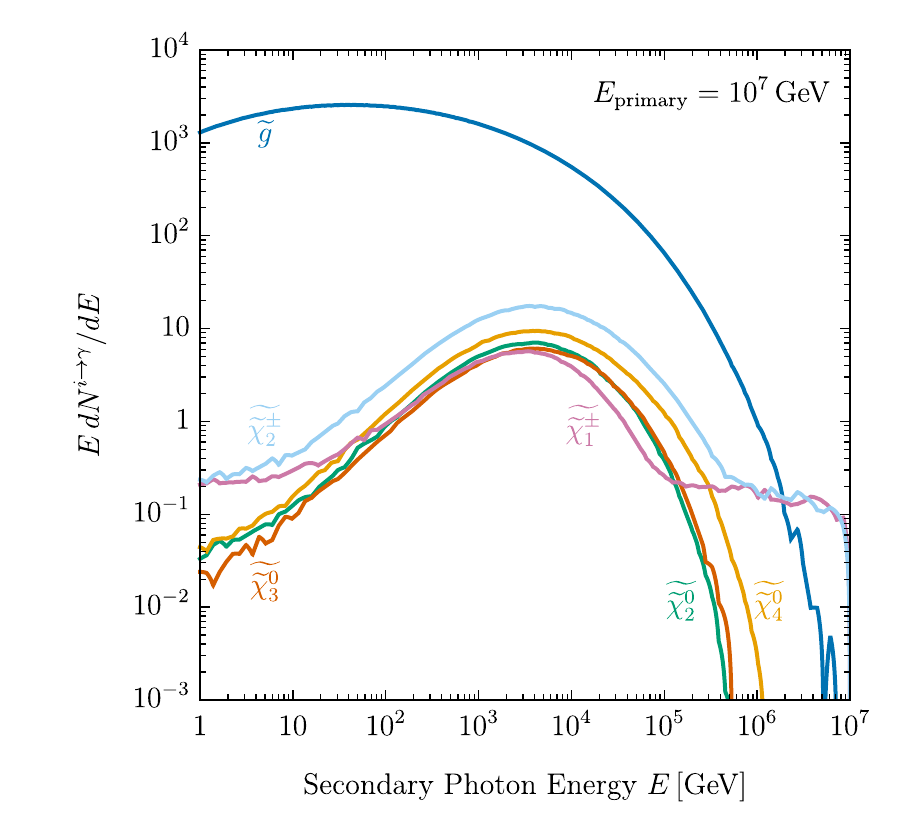}&
    \includegraphics[width=0.475\textwidth]{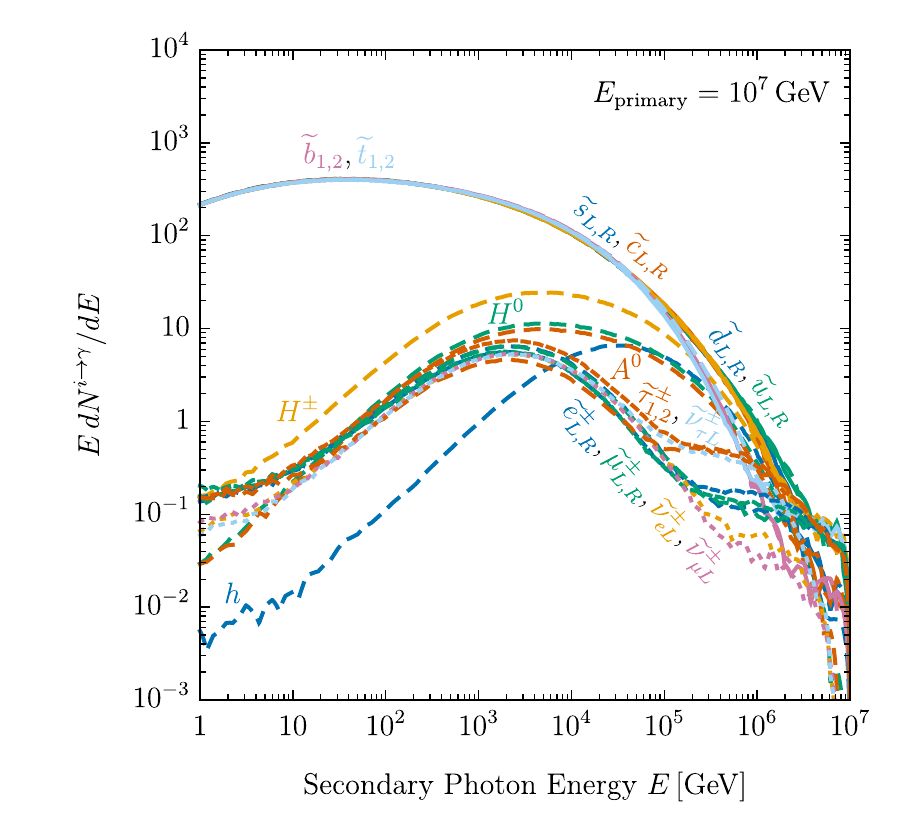}
    \end{tabular}
    \caption{The secondary photon spectra per primary particle for the MSSM fermions (left) and bosons (right) at a primary energy of $10^4\,\text{GeV}$ (top) and $10^7\,\text{GeV}$ (bottom).  The artefacts are Monte Carlo error for rare processes and do not lead to a significant error in the total secondary spectra.}
    \label{fig:mssm-secondaries}
\end{figure}
%-----------------------------------------------------------------

In \cref{fig:mssm-secondaries} we present the secondary photon spectra for the MSSM particles, in analogy to those for the SM particles shown in \cref{fig:secondaries}.  In the top left panel of \cref{fig:mssm-secondaries} we show the secondary spectra for the MSSM fermions at a primary energy of $10^4\,\text{GeV}$.  The gluino, since it is coloured and has 16 degrees of freedom, makes the largest contribution.  The neutralinos make a significantly smaller contribution, which is largest at low energies.  The contribution from the charged neutralinos, $\widetilde\chi_1^\pm$ and $\widetilde\chi_2^\pm$, stops falling off with higher energy and becomes the dominant contribution, even dominating over the gluino, at the highest energies.  While $\widetilde\chi_1^\pm$ predominantly decays to $\widetilde\chi_1^0 \bar f f'$, where $f$ and $f'$ are SM fermions, $\widetilde\chi_2^\pm$ dominantly decays to $\widetilde\chi_1^0 W^\pm$ and $\widetilde\chi_1^\pm Z$, leading to a higher photon flux at low and mid energies.  The LSP, $\widetilde{\chi}_1^0$, essentially does not produce any secondary photons.  As in \cref{fig:secondaries}, Monte-Carlo error is seen when the rates are small but this does not significantly impact the integrated secondary spectra, \cref{eq:d2NdEdt-secondaries}.

In \cref{fig:mssm-secondaries} (top right) we show the secondary photon spectra for the MSSM bosons of an energy of $10^4\,\text{GeV}$.  We also include the SM-like Higgs boson since its decays are slightly different to the SM scenario.  We see that at low energies, the largest contributors are the squarks, followed by the heavier Higgs bosons, $H^0$, $A^0$ and $H^\pm$, then the sleptons and finally the lightest Higgs boson, $h$.  At mid-energies the lightest Higgs boson, $h$, becomes as important as the other Higgs bosons, and the staus and tau-sneutrinos become more important than the other sleptons.

Similar behaviour is seen at $10^7\,\text{GeV}$, shown in the lower panels of \cref{fig:mssm-secondaries}.  As in the SM, the peaks are shifted to higher energies due to the higher energy of the primary particles.  

The supersymmetric particles have broadly similar secondary spectra to their SM counterparts, with the exception of the sleptons and leptons. This is because the sleptons readily decay to coloured SM particles via neutralinos and charginos, which go on to produce many photons.

%=============================================================================
\bibliographystyle{JHEP}
\bibliography{refs}
%=============================================================================

\end{document}